\documentclass{aa}
\usepackage{color}
\usepackage[colorinlistoftodos]{todonotes}
\usepackage{txfonts,bm}
\usepackage{amssymb}
\usepackage{graphicx}

\newcommand{\HI}{{\ion{H}{1}}}

\newcommand{\WHz}{$\,$W$\,$Hz$^{-1}$}

\newcommand{\mJybeam}{mJy beam$^{-1}$}
\newcommand{\muJybeam}{$\mu$Jy beam$^{-1}$}

\newcommand{\si}{$\alpha^{\rm 150}_{\rm 1400}$}

\def\HI{\ion{H}{i}}

\def\emph#1{{\sl #1}}
\newcommand{\ltsima} {$\; \buildrel < \over \sim \;$}
\newcommand{\gtsima} {$\; \buildrel > \over \sim \;$}
\newcommand{\lta} {\lower.5ex\hbox{\ltsima}}
\newcommand{\gta} {\lower.5ex\hbox{\gtsima}}

\begin{document}

\title{The best of both worlds:
Combining LOFAR and Apertif \\ 
to derive resolved radio spectral index images }


\author{
R. Morganti\inst{1,2},
T.A. Oosterloo\inst{1,2}
M. Brienza\inst{3,4}, N. Jurlin\inst{1,2},
I. Prandoni\inst{4}, E. Orru`\inst{1}, S.S. Shabala\inst{5,6},
E.A.K.~Adams$^{1,2}$, B.~Adebahr$^{7}$, P.N. Best$^{8}$
A.H.W.M.~Coolen$^{1}$, S.~Damstra$^{1}$, W.J.G.~de~Blok$^{1,2,9}$, F. de Gasperin\inst{10},\\ H. D\'enes$^{1}$, M. Hardcastle\inst{11}, K.M.~Hess$^{1,2}$, B.Hut$^{1}$, R. Kondapally\inst{8}, A.M. Kutkin$^{1,12}$, G.M.~Loose$^{1}$, D.M.~Lucero$^{2,13}$, Y.~Maan$^{14}$, F.M. Maccagni$^{15}$, B.Mingo\inst{16}, V.A. Moss$^{1,17,18}$, R.I.J. Mostert$^{1,19}$, M.J.~Norden$^{1}$,  L.C.~Oostrum$^{1,14}$, \\   H.J.A. R\"ottgering\inst{19}, 
M.~Ruiter$^{1}$,
T.W. Shimwell$^{1,19}$, 
R.~Schulz$^{1}$,
N.J.~Vermaas$^{1}$,
D.~Vohl$^{1}$, 
J.M.~van~der~Hulst$^{2}$, \\
G.M. van Diepen$^{1}$,
J.~van~Leeuwen$^{1,14}$,
J.~Ziemke$^{1,20}$
}

\offprints 
{Raffaella Morganti \\
\email{morganti@astron.nl}}

\authorrunning{Morganti et al.}
\titlerunning{Resolved spectral index images with Apertif and LOFAR}

\institute{ASTRON, the Netherlands Institute for Radio Astronomy, Oude Hoogeveensedijk 4, 7991 PD Dwingeloo, The Netherlands
\and
Kapteyn Astronomical Institute, University of Groningen, Postbus 800, 9700 AV Groningen, The Netherlands
\and 
    Dipartimento di Fisica e Astronomia, Universit\`a di Bologna, Via P.\ Gobetti 93/2,     I-40129, Bologna, Italy
\and 
    INAF - Istituto di Radio Astronomia, Via P.\ Gobetti 101, I-40129 Bologna, Italy
\and
   School of Natural Sciences, Private Bag 37, University of Tasmania, Hobart, TAS 7001, Australia.
\and
    ARC Centre of Excellence for All-Sky Astrophysics in 3 Dimensions (ASTRO 3D).
\and
    Astronomisches Institut der Ruhr-Universit{\"a}t Bochum (AIRUB), Universit{\"a}tsstrasse 150, 44780 Bochum, Germany 
\and 
    SUPA, Institute for Astronomy, Royal Observatory, Blackford Hill, Edinburgh, EH9 3HJ, UK 
\and 
    Department of Astronomy, University of Cape Town, Private Bag X3, Rondebosch 7701, South Africa
\and Hamburger Sternwarte, Universit\"a Hamburg, Gojenbergsweg 112, 21029, Hamburg, Germany
\and
    Centre for Astrophysics Research, School of Physics, Astronomy and Mathematics, University of Hertfordshire, College Lane, Hatfield, Hertfordshire, AL10 9AB, UK.
\and
    Astro Space Center of Lebedev Physical Institute, Profsoyuznaya Str.\ 84/32, 117997 Moscow, Russia
\and Department of Physics, Virginia Polytechnic Institute and State University, 50 West Campus Drive, Blacksburg, VA 24061, USA 
\and Anton Pannekoek Institute, University of Amsterdam, Postbus 94249, 1090 GE Amsterdam, The Netherlands
\and INAF - Osservatorio Astronomico di Cagliari, via della Scienza 5, 09047, Selargius (CA), Italy
\and School of Physical Sciences, The Open University, Walton Hall, Milton Keynes, MK7 6AA, UK 
\and 
CSIRO Astronomy and Space Science, Australia Telescope National Facility, PO Box 76, Epping NSW 1710, Australia
\and Sydney Institute for Astronomy, School of Physics, University of Sydney, Sydney NSW 2006, Australia
\and Leiden Observatory, Leiden University, Postbus 9513, 2300 RA Leiden, The Netherlands. 
\and Center for Information Technology, University of Groningen, Postbus 11044, 9700 CA Groningen, the Netherlands
   }

\date{version \today}

\abstract
{Supermassive black holes at the centres of galaxies can cycle through periods of activity  and quiescence. Characterising the duty cycle of active galactic nuclei (AGN) is crucial for understanding the impact of the energy they release on the host galaxy.  For radio AGN, this can be done by identifying dying (remnant) and restarted radio galaxies from their radio spectral properties.
Using the combination of the images at 1400~MHz produced by Apertif, the new phased-array feed receiver installed on the Westerbork Synthesis Radio Telescope, and images at 150 MHz provided by LOFAR, we have derived resolved spectral index images (at a resolution of $\sim$15 arcsec) for all the sources within an approximately 6 deg$^2$ area of the Lockman Hole region. In this way, we were able to select 15 extended radio sources with emission (partly or entirely) characterised by extremely steep spectral indices (steeper than 1.2). These objects represent cases of radio sources in the remnant or the restarted phases of their life cycle.
Our findings confirm that these objects are not as rare as previously thought, suggesting a relatively fast cycle. They also show a variety of properties that can be relevant for  modelling the evolution of radio galaxies. For example, the restarted activity can occur while the remnant structure from a previous phase of activity is still visible. This provides constraints on the duration of the 'off' (dying) phase.  In extended remnants with ultra-steep spectra  at low frequencies, the  activity likely stopped a few hundred megayears ago, and they correspond to the older tail of the age distribution of radio galaxies, in agreement with the results of  simulations of radio source evolution.  We find remnant radio sources with a variety of structures (from double-lobed to amorphous), possibly suggesting different types of progenitors. 
The present work sets the stage for exploiting the powerful tool of low-frequency spectral index studies of extended sources by taking advantage of the  large areas common to the LOFAR and the Apertif surveys. }


\maketitle

\section{Introduction}
\label{sec:introduction}

The recurrence of the active phase of the super massive black hole (SMBH)  in the centre of galaxies is a key ingredient in cosmological simulations. It ensures that the energy released has an impact on the environment multiple times during the life of the host galaxy. In the feedback scenario for galaxy evolution, this is necessary in order to prevent the cooling of the gas from their hot halo onto massive galaxies  and to regulate and quench their star formation (see e.g.\ \citealt{Ciotti10,Hobbs11,Novak11,Heckman14,Gabor13,Gaspari17}). For radio active galactic nuclei (AGN), the advent of a new generation of low-frequency surveys with high spatial resolution has opened up the possibility of making substantial progress in quantifying these cycles of activity.
This is because the low-frequency emission ($<$1400 MHz) is  the least affected by the radiative losses occurring in the sources and, therefore, can act as a `fossil record', tracing the history of the radio emission over long timescales (e.g.\ \citealt{Morganti17}).

It has been known for a long time that in radio AGN a period of activity ends with a remnant phase (often characterised by an amorphous, low surface brightness structure) when the nuclear activity switches off or dims drastically, and that this can be followed by a subsequent period of activity.
However, quantifying the fraction of these remnant and restarted radio sources and timing the cycle of activity has been hindered by the difficulty of finding such sources. 
 They have mostly been discovered  serendipitously  (e.g.\ \citealt{Cordey87,Shulevski17,Brocksopp07,Orru15,Kuzmicz17,Bruni20} for some examples) or by targeting sources with specific morphological properties like the so-called double-double sources (e.g.\ \citealt{Schoenmakers00,Konar13,Kuzmicz17}) or sources with extremely steep integrated spectral indices (e.g.\ \citealt{Parma07,Murgia11}). All together, these studies have only provided a few dozen cases, selected in an heterogenous way. The first attempt to perform a blind morphological selection was presented by \cite{Saripalli12}.
 
Building large and representative samples of radio sources that are in these elusive phases has now become possible thanks to imaging surveys with radio telescopes that have a large field of view at low frequencies (e.g.\ the Low Frequency Array (LOFAR) Two-metre Sky Survey  (LoTSS), \citealt{Shimwell19}; the GaLactic and Extra-Galactic All-Sky MWA  (GLEAM) survey, \citealt{Hurley17}; and the  Tata Institute of Fundamental Research (TIFR) Giant Metrewave Radio Telescope (GMRT) Sky Survey (TGSS), \citealt{Intema17}). 
From the studies based on these surveys, once combined with appropriate modelling, a picture starts to emerge regarding the details of the cycle of activity of radio AGN (see \citealt{Brienza17,Godfrey17,Hardcastle18,English19,Mahatma19,Shabala20}). 

For example, these studies have shown that, for  a large fraction of remnant sources found based on their morphology, the central source must have switched off recently and that remnant sources should fade relatively fast (on timescales of at most a few $10^8$ yr) due to  both radiative losses and dynamical expansion (\citealt{Brienza17,Godfrey17,Hardcastle18,Turner18}). 
In addition, the recent finding of a significant number of candidate restarted radio sources (again, mostly from a morphological selection, see \citealt{Jurlin20} for details) has led to the suggestion that a new phase of activity can start relatively quickly after the switching-off of the previous phase. All this requires a dominant population of short-lived jets, as confirmed by the detailed modelling presented by \cite{Shabala20}. This cycle of `on' and `off' activity can become even more rapid for low power radio sources  in massive early-type galaxies \citep{Sabater19} and for radio sources in particularly gas-rich clusters (i.e.\ cool-core clusters, e.g.\ \citealt{McNamara12,Vantyghem14}).

These obtained results have so far been mostly based on morphological signatures observed at low radio frequencies or on the properties of the integrated radio spectrum. 
However, the cycles should also leave signatures in the spectral distribution across the radio source, making the study of the  resolved spectral properties an interesting tool for complementing and expanding our understanding of radio galaxy evolution.

Since the development of the theory of synchrotron emission, the spectral index\footnote{In this paper, the spectral index $\alpha$ is defined through $S\propto \nu^{-\alpha}$.} and the shape of the spectrum  of the radio emission have been used to characterise the evolution of the radio plasma, allowing us to trace radiative losses and locate regions of particle re-acceleration (see e.g.\  \citealt{Kellerman64,Kardashev62,Pacholczyk70,Jaffe73,Komissarov94}). These spectral properties have been  used extensively for doing `archaeology' on radio galaxies (see e.g.\ \citealt{Morganti17,Hardcastle20} for reviews), albeit with a number of limitations that should be kept in mind. These include both technical considerations (e.g.\ matching resolution) and physical effects (e.g.\ structure of the magnetic field, particle re-acceleration, and particle mixing, see \citealt{Turner18} and \citealt{Eilek96} for a discussion).

Particularly relevant for this study are ultra-steep spectrum (USS) indices (steeper than $\sim$1.2; \citealt{Komissarov94}), which suggest that the replenishment of fresh electrons has stopped and the spectrum is dominated by losses (from radiation and expansion), properties characteristic of sources in the dying phase (e.g.\ \citealt{Parma07}). This steepening is more often seen at high frequencies (i.e.\  $> 1400$ MHz) because of the higher radiative  losses of higher-energy electrons. 
The low-frequency spectrum ($<$1400 MHz) is instead the last affected by  the radiative losses of the source. The radio spectra at these frequencies can maintain the original injection index of the electrons (typically between 0.5 -- 0.8, see e.g.\ \citealt{Harwood16}) for longer.  Thus, the presence of  steepening in the spectrum and the frequency above which it occurs are, to the first order,  functions of the time passed since the last re-acceleration or injection of fresh electrons. 
Finding spectral indices steeper than $\sim$1 at low frequencies is a strong indication that the impact of energy losses has reached those frequencies, suggesting aged remnant plasma and thus a source where the nuclear activity stopped long ago. 

This technique is even more powerful when it can be traced and resolved across the emission of the source. This gives a view of the history of the source and can allow us to separate various phases of activity and time them.  This has been done for a number of sources at high frequencies, while a more limited number of studies is available at low frequencies. Some examples include: active Fanaroff-Riley \citep{Fanaroff74} radio galaxies type I and II (FRI and FRII; e.g.\ \citealt{Orru10,Heesen18,Harwood16}); double-double radio galaxies; and a few famous cases of restarted sources (see e.g.\ \citealt{Konar13,Jamrozy07,Joshi11,Brienza20}) and  remnant radio sources (e.g.\ \citealt{Murgia11,Shulevski15,Shulevski17,Brienza16,Murgia11,Gasperin14,Hurley15,Tamhane15}). Even rarer are the objects where the presence of two episodes of jet activity can only be seen from the discontinuity in the spectral index distribution (see e.g.\ 3C~388, \citealt{Roettiger94,Brienza20}).
  Thus,  a systematic study of resolved low-frequency radio spectra for large samples of radio galaxies is greatly needed in order to confirm and expand the picture we are assembling of the evolution of radio sources.

\section{New possibilities for spectral index studies over large areas}

The advent of LOFAR (\citealt{Haarlem13}) and the excellent spatial resolution it provides opens up a number of opportunities for spectral index studies down to megahertz frequencies. However,  resolved spectral index studies have  so far been limited to single objects or small sky areas. This is due to the lack of ancillary data at other radio frequencies with comparable spatial resolution and depth.  Only for samples of compact sources have the integrated spectral properties been derived using a combination of data -- with different spatial resolutions -- from available public surveys (e.g.\ the NRAO Very Large Array (VLA) Sky Survey (NVSS), Westerbork Northern Sky Survey (WENSS), TGSS, etc.). 
From these studies, a median spectral index of \si$= 0.78 \pm 0.015$ has been found for active sources (see e.g.\  \citealt{Intema11,Weeren14,Mahony16,Williams16}). 
However, even these studies are limited by the sensitivity of the available surveys, which restricts the number of objects that can be studied. 

The launch of imaging surveys with the APERture Tile In Focus (Apertif) phased-array feed (PAF) system, recently installed on the Westerbork Synthesis Radio Telescope (WSRT) and working at 1400 MHz (see \citealt{Oosterloo18,Adams19} for more details), has provided  an ideal complement to the surveys done with  LOFAR  at frequencies centred on 150 and 54 MHz (\citealt{Shimwell19} and de Gasperin et al. in preparation).
The main asset of Apertif is its combination of a large field of view (roughly $3\fdg5\times3^\circ$) and relatively high spatial resolution ($\sim$12 arcseconds).

Here we take advantage of the combination of the continuum images provided by LOFAR at 150 MHz and Apertif at 1400 MHz  and present a proof-of-concept study of the resolved spectral index of extended radio sources in the Lockman Hole (LH) area. 
This is the first time spectral index maps have been produced blindly on a large piece of sky at high spatial resolution. The only previous attempt in this direction was the work by \cite{Gasperin18}, who performed a joint extraction from TGSS and NVSS. That study is, however, limited by the spatial resolution of the NVSS and the sensitivity of the TGSS.

Our study does not aim to carry out a full characterisation of the spectral properties. Instead,  it focuses on the search for extended radio sources with the resolved spectral properties expected for objects in the remnant and restarted AGN phases. Based on the arguments described above,  we look for radio sources where (part of) the emission is characterised by a  USS, suggesting the presence of  aged plasma, thus giving information on their past evolution. 

The LH  area is particularly interesting for this because we have started, using  the LOFAR 150~MHz observations of this field, to identify radio galaxies that are in the remnant and restarted phases and to derive the statistics of the fraction of sources in these elusive phases. The results so far have been presented in a series of papers: \citet{Brienza17}, \citet{Jurlin20}, and \citet{Shabala20}. 
The selection of remnant and restarted radio sources in these papers has mainly relied on the radio morphology and, in some cases, on their integrated spectrum. For the reasons described above, this now needs to be complemented using the resolved spectral properties of the sources in the field. 
This is the goal of the present study.

The paper is organised as follows. 
In Sect. \ref{sec:observations}, we describe the observations obtained with Apertif during its commissioning phase,  we recap the LOFAR observations that will be used, and we discuss how the spectral indices were derived.
In Sect.\ \ref{sec:results}, we describe the spectral properties of   two groups of sources (Sects. \ref{sec:partlyUSS} and \ref{sec:USS}) that were identified and extracted. We also briefly remark in Sect.\  \ref{sec:A1132} on the spectral properties of some of the sources in the cluster Abell~1132, which is included in our field.  In Sect.\ \ref{sec:statistics}, we discuss the results in relation to the sample used by \cite{Jurlin20} to derive the occurrence of remnant and restarted radio sources. In Sect.\  \ref{sec:relevance}, the relevance of our findings is discussed in the context of understanding the life cycle of radio galaxies. Conclusions and our future outlook are given in Sect.\ \ref{sec:conclusions}.

\begin{figure*}
   \centering
\includegraphics[width=15cm,angle=0]{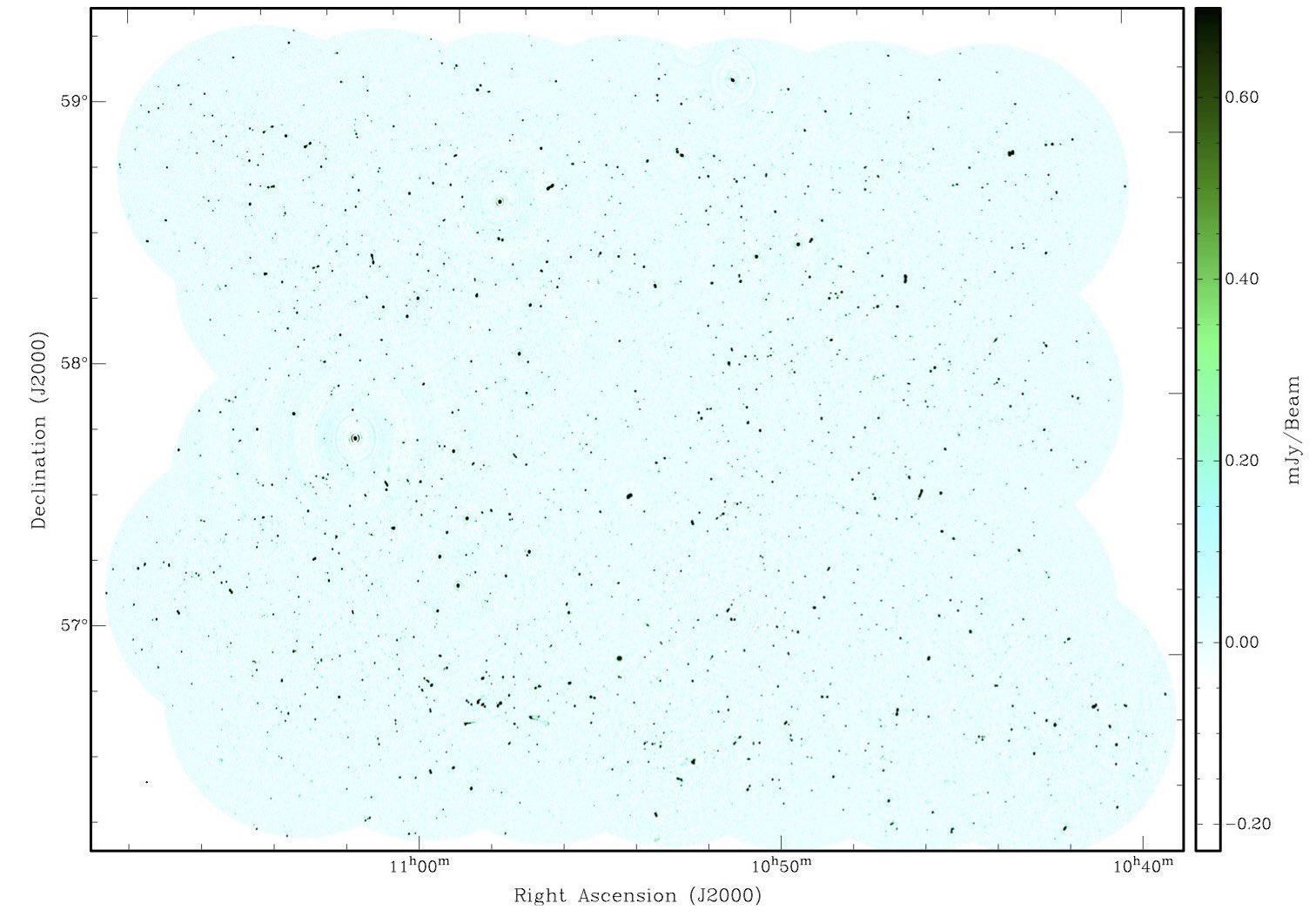}
   \caption{Image obtained at 1400~MHz from the Apertif observations. The image is the result of the mosaicing of the 40 Apertif beams. The large complex of sources that can be seen at the bottom of the image -- around the location  RA=$10^{\rm h}58^{\rm m}00^{\rm s}$  Dec=$+56^\circ46^\prime00^{\prime\prime}$ (J2000) -- is the cluster Abell~1132. }
\label{fig:ApertifField}
\end{figure*}

\begin{figure}
   \centering
\includegraphics[width=9cm,angle=0]{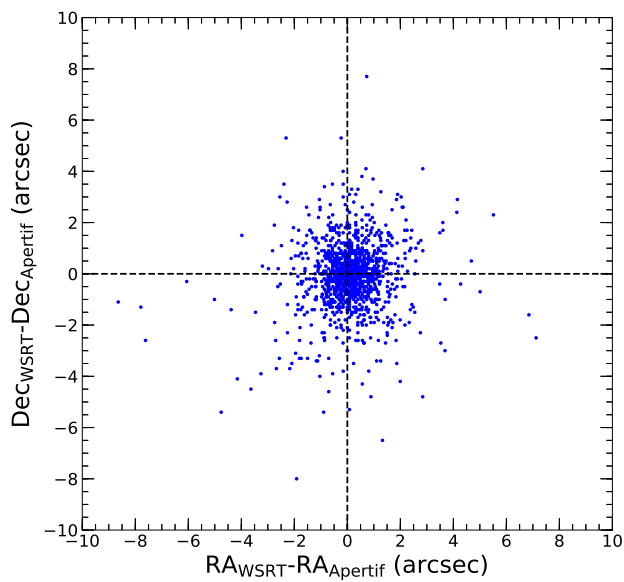}
   \caption{Distribution of the position offsets between sources in the old WSRT image and in the Apertif mosaic after correcting for the median offset (see text for details).}
\label{fig:positionsComparison}
\end{figure}

\begin{figure}
   \centering
\includegraphics[width=9cm,angle=0]{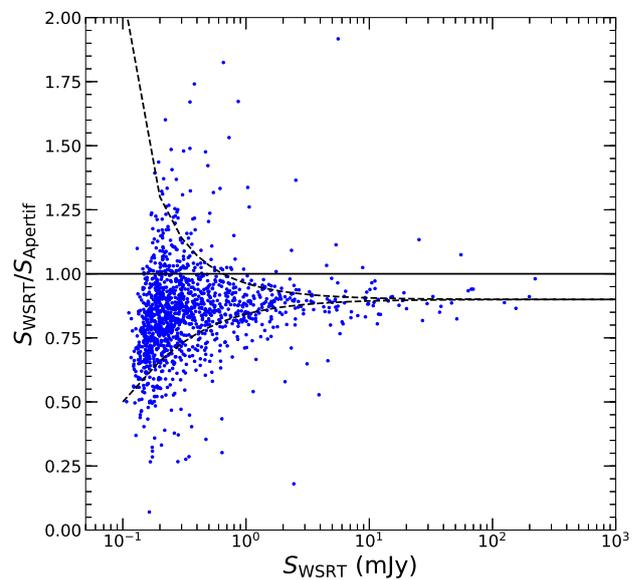}
   \caption{Distribution of the flux ratio of sources in the old WSRT image and the new Apertif image. An offset of 10\% in the Apertif flux scale is indicated and has been corrected for in the analysis (see text for details). The dashed lines indicate the expected spread due to an error in the flux densities of 1$\sigma$. }
\label{fig:fluxesComparison}
\end{figure}

\section{The Lockman Hole region observed  by Apertif and LOFAR
}
\label{sec:observations}

Apertif is the  phased-array  system recently installed at the WSRT.  Using the PAFs in the focus of  each WSRT dish, 40 partially overlapping beams, each having a full width at half maximum (FWHM) of about 35 arcminutes, are formed on the sky. The data from each beam are  used independently to image, using standard aperture synthesis, the region covered by each beam. After this, these 40 images are stitched  together to form the final mosaiced image. The total area covered by such a mosaic is about $3\fdg5\times3^\circ$, and the  spatial resolution  is about $12^{\prime\prime} \times 16^{\prime\prime}$ at the declination of the LH (see \citealt{Oosterloo18,Adams19} for more details on Apertif and van Cappellen et al.\ in preparation\ for a technical description of the system). The Apertif Imaging Surveys started on July 1, 2019. A description of these surveys is given by Hess et al.\ (in preparation), but here we use data taken earlier during the science commissioning phase. 

The observations of the LH were done on April 28, 2019, during the Apertif commissioning phase. The target field was observed for 11 h, while the flux and bandpass calibrator  (3C~147) was observed at the start of the observations for 3 min for each of the 40 Apertif beams. The useful bandwidth, after flagging  a substantial part of the observing band  because of radio frequency interference, was 200 MHz centred on 1408 MHz, with a spectral resolution of 0.78125 MHz.

The Apertif observation was centred on RA=$10^{\rm h}54^{\rm m}00^{\rm s}$  Dec=$+57^\circ50^\prime00^{\prime\prime}$ (J2000). This position was chosen to provide  maximum overlap with the mosaic of the LH made at the same frequency with the old WSRT system \citep{Prandoni18}.

 These previous observations covered an area of about 6 deg$^2$, but only the inner $\sim$2 deg$^2$ region had a uniform noise of $\sim 11$ \muJybeam (and rapidly increasing towards the edges, see \citealt{Prandoni18} for details). Covering this area  required 16 observations of 12 h each. In contrast, the Apertif observation has allowed us to cover about 10 deg$^2$ in a single 12-h observation, albeit with slightly higher noise (due to the  PAF receivers being uncooled, unlike the old WSRT receivers). This resulted in an image  of about 6 deg$^2$  with uniform noise, greatly expanding the area usable for deriving spectral indices and for searching for remnant and restarted radio galaxies.
 In practice, the survey speed of Apertif is about a factor of ten higher than that of the old WSRT system, showing the potential for covering large areas in a limited observing time. 
However, the well-characterised deep WSRT mosaic still provides a reference point thanks to the  accuracy  of  its  astrometry  and   flux  scale (as described by Prandoni et al.\ 2018), allowing a detailed check of the Apertif source positions and flux densities (see below), which is appropriate considering that these data were taken in the commissioning phase.

The flagging and cross-calibration of the Apertif data were done using an early version of Apercal, the Apertif pipeline (Schulz et al.\ 2019, Adebahr et al.\ 2020 in preparation). 
Automated flagging is performed inside this pipeline by the AOFlagger software \citep{Offringa10,Offringa12}.
The subsequent reduction steps (i.e.\ self-calibration, imaging, cleaning, and mosaicing) and the analysis were done using scripts that made use of the Miriad package (\citealt{Sault95}). Briggs weighting \citep{Briggs95} was used (robust $=-1.5$) to make the images. To correct for small differences in resolution between the images from the different Apertif beams, the cleaned images were convolved (before mosaicing) to a common resolution of  $16.5 \times 12.5$ arcsec (in PA$=-0.1^\circ$). It is important to note that the array configuration of the WSRT includes short baselines down to 36 m. This ensures sensitivity to structures up to at least 10 arcmin, as demonstrated by many results obtained in the past by the WSRT telescope  (see e.g. \citealt{Barthel85,Oosterloo07,Shulevski12}).

Given that Apertif was still in the commissioning phase, a number of the images  suffered from direction-dependent errors due to small errors in the weights applied to the PAF elements used in forming the 40 Apertif beams.  These  direction-dependent errors were corrected for using a standard peeling algorithm implemented in scripts based on Miriad. Nevertheless, some artefacts remained around some bright sources.  The resulting final Apertif mosaic is shown in Fig.\ \ref{fig:ApertifField}. The rms noise reached in this image is $\sim$30 \muJybeam\ over most of the image and increasing towards the edges of the mosaic. The large complex of sources that can be seen at the bottom of the image is the cluster  Abell 1132 (see e.g. \citealt{Wilber18}).

The LOFAR image at 150~MHz was obtained from 12 observations of $\sim$8 hrs (100 hrs total integration time) with the same pointing centre, as described in Tasse et al.\ (2020) and is being released as part of the LoTSS Deep Fields Data Release 1. This image has been used for the selection of restarted radio sources and comparison samples as described by \cite{Jurlin20}. This image was obtained using direction-dependent calibration (see \citealt{Shimwell19} and  \citealt{Tasse20}). The spatial resolution of the  image is $6$ arcsec,  the maximum resolution that can be obtained at 150~MHz with the Dutch stations of LOFAR. The area covered by a single LOFAR pointing is much larger than the Apertif mosaic: The common area is about 2\fdg5 $\times$ 2\fdg5  after excluding the edge of the Apertif mosaic, that has higher noise. In order to match the spatial resolution so to be able to derive spectral indices, the LOFAR image was convolved to exactly the same resolution as that of the Apertif mosaic. The resulting lower-resolution LOFAR image has an rms noise of $\sim$ 0.11 \mJybeam,  higher, as expected, than in the original higher-resolution image (see  \citealt{Tasse20} and  \citealt{Sabater20} for details). The relative 1$\sigma$ noise levels of the Apertif and LOFAR images  correspond to a spectral index of $\alpha = 0.58$.

As mentioned above, part of the area covered by Apertif was observed earlier at 1400 MHz using the old WSRT receiver system  \citep{Prandoni18}.
Because a proper alignment of the positions and of the flux scale is important for making spectral images, we used this earlier mosaic to check the accuracy of the  fluxes and positions of the sources extracted from the Apertif image before producing images of the spectral index.

An offset of about 2\farcs5  in right ascension and 1\farcs3  in declination was found between the position of the sources in the Apertif image and in the old WSRT mosaic. This is not  unexpected because  a close astrometric calibrator (that was observed regularly during the target observations) was used for the old WSRT mosaic, while this was not the case for the Apertif observations due to time constraints in the observing schedule. A comparison with the source positions in the LOFAR image confirms this offset and shows that it is of consistent amplitude across the entire field. The Apertif images were corrected for this shift before mosaicing. The comparison of the positions of sources common to the Apertif and the old WSRT images, after this correction,  is shown in Fig.\  \ref{fig:positionsComparison}. 

An offset   was  also found for the  flux densities, with  the Apertif fluxes being about 10\% higher,   as illustrated in Fig.\ \ref{fig:fluxesComparison}. This is likely due to the fact that, at the time of the observations, the Apertif calibration algorithms had not yet been fully developed.  The fluxes in the Apertif image were corrected for this offset in the flux scale. The rms noise used for the spectral images was derived on the scaled image. A detailed discussion of the flux scale of the LOFAR image is presented by   \cite{Tasse20}. 

\begin{figure*}
   \centering
\includegraphics[width=6.4cm,angle=-90]{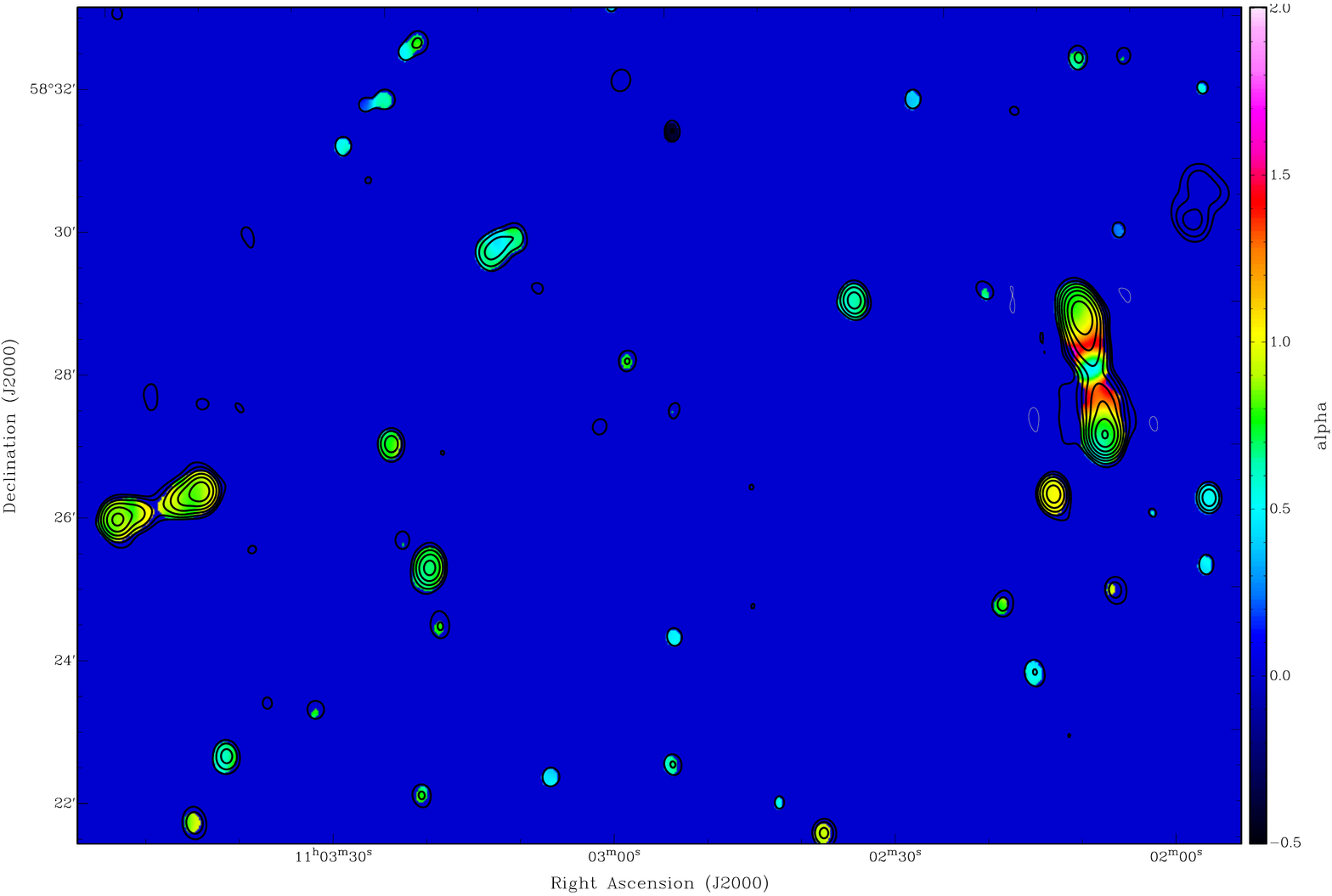} 
\includegraphics[width=6.4cm,angle=-90]{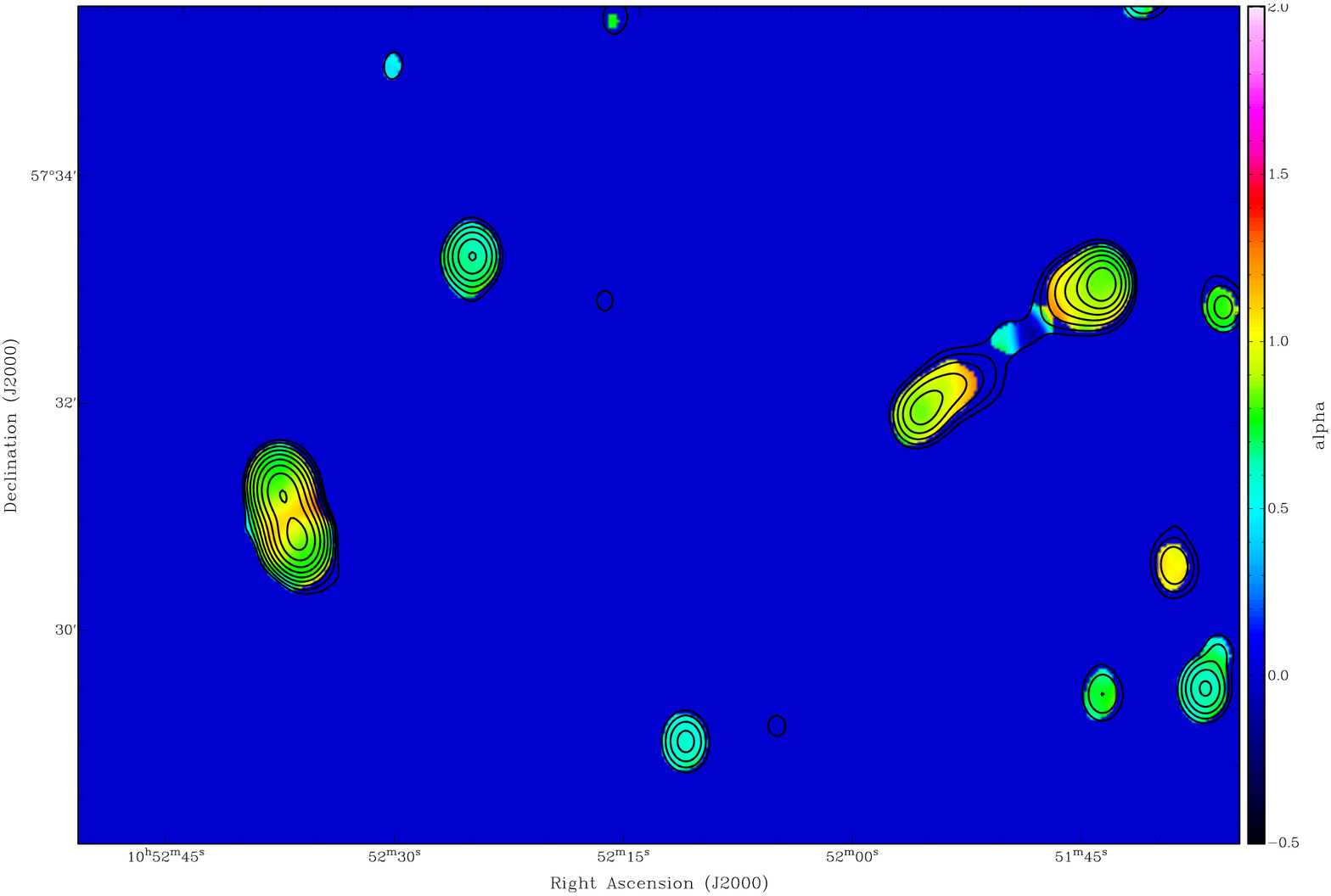} \\
\includegraphics[width=6.4cm,angle=-90]{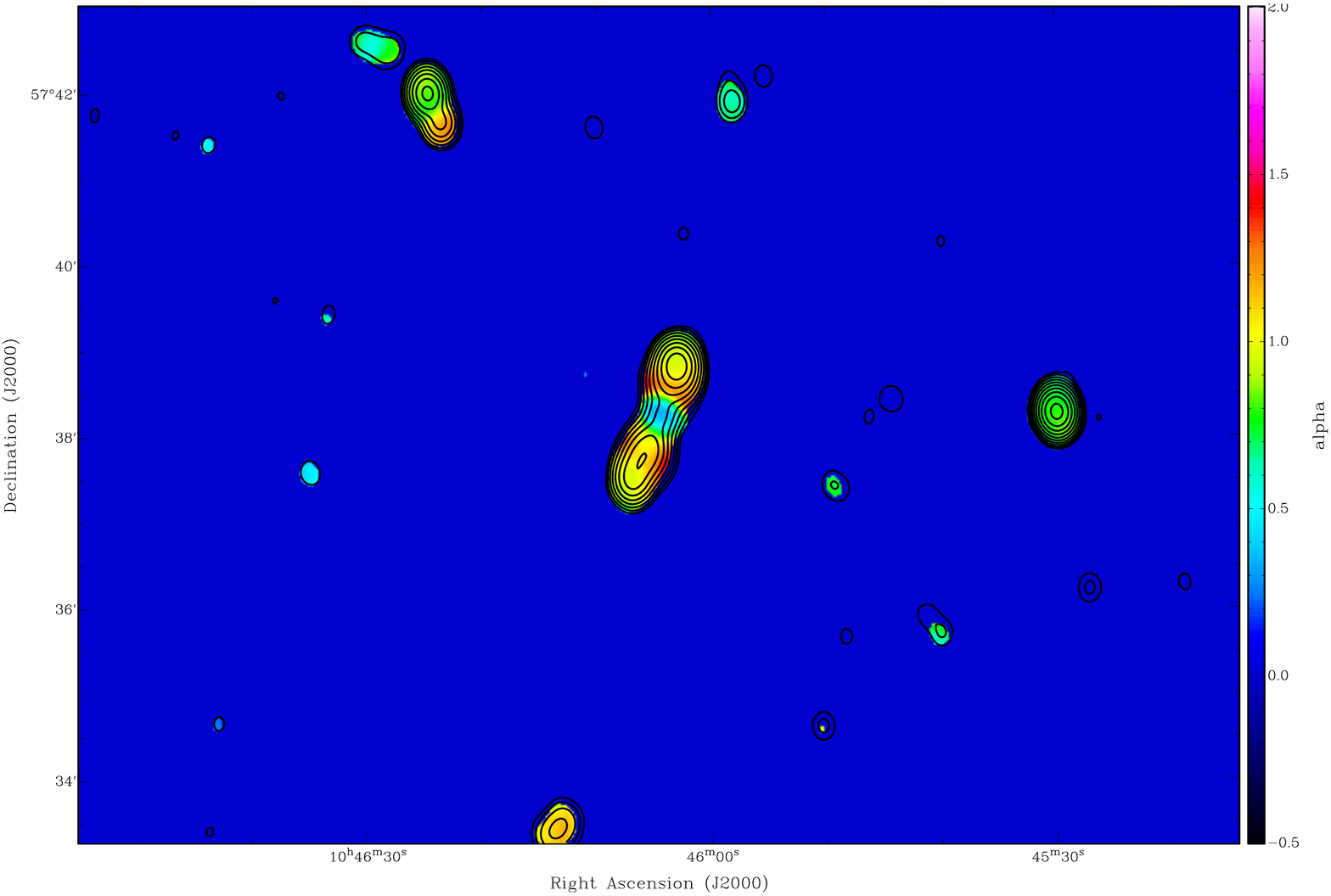} 
\includegraphics[width=6.4cm,angle=-90]{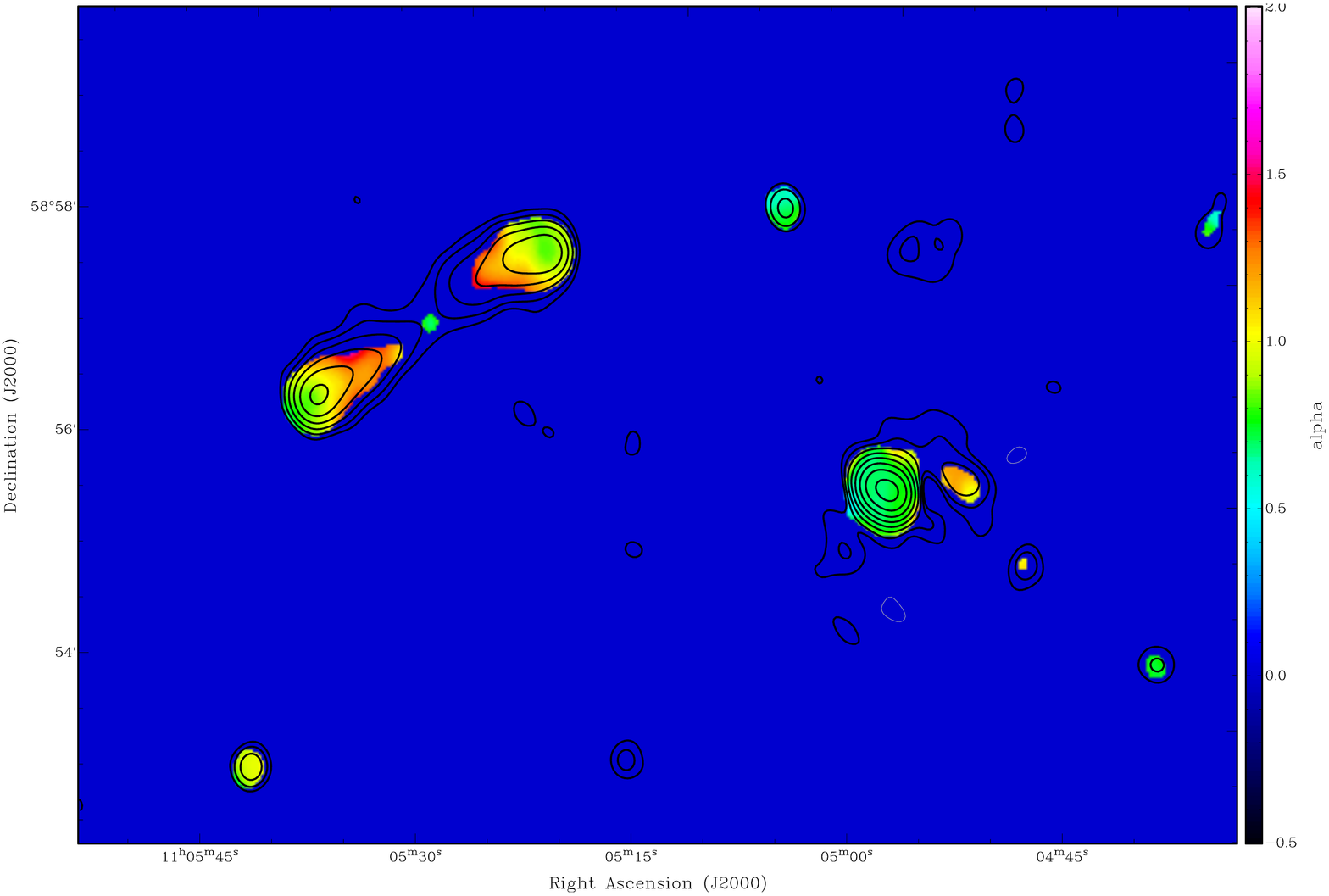} \\
\includegraphics[width=6.4cm,angle=-90]{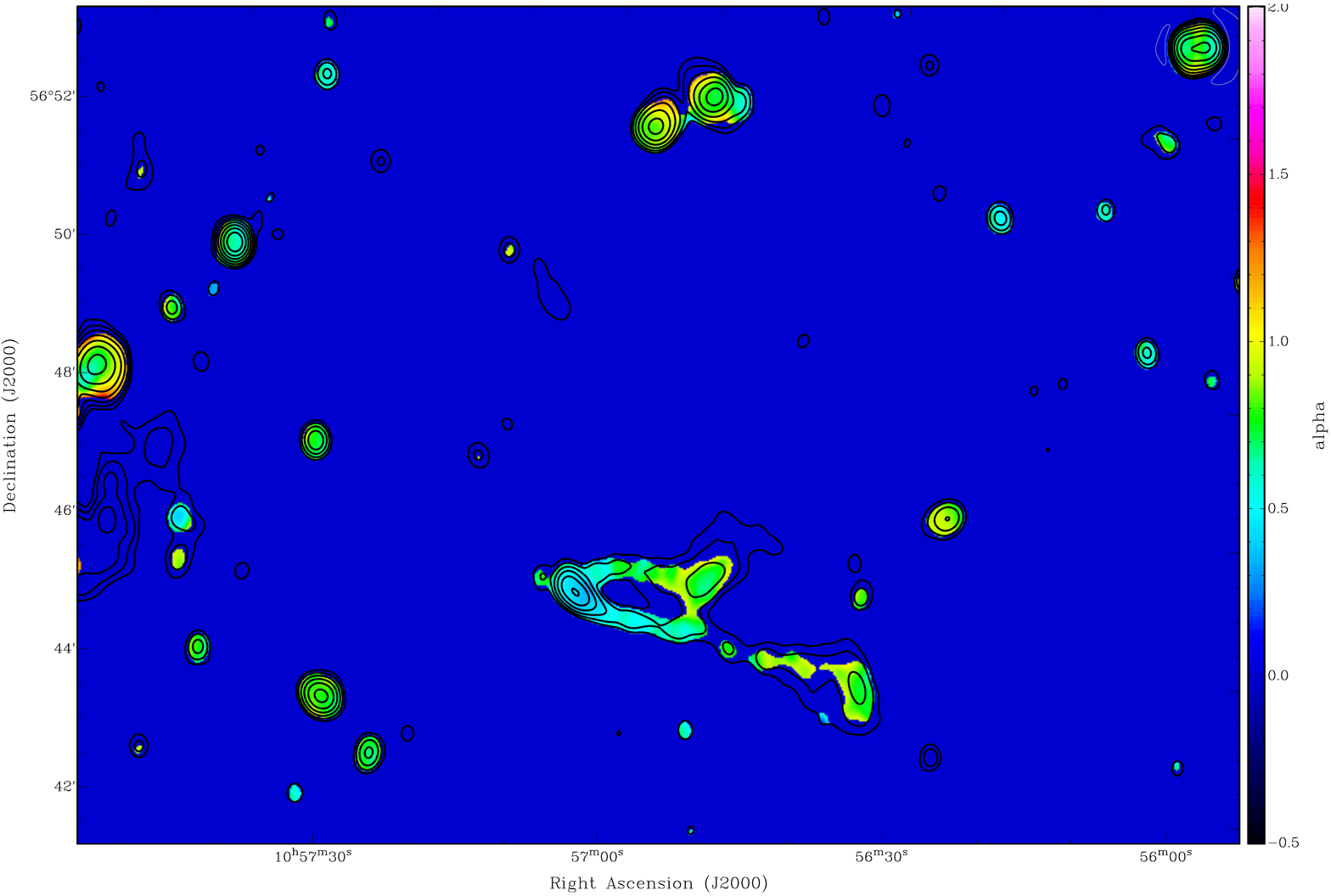} 
\includegraphics[width=6.4cm,angle=-90]{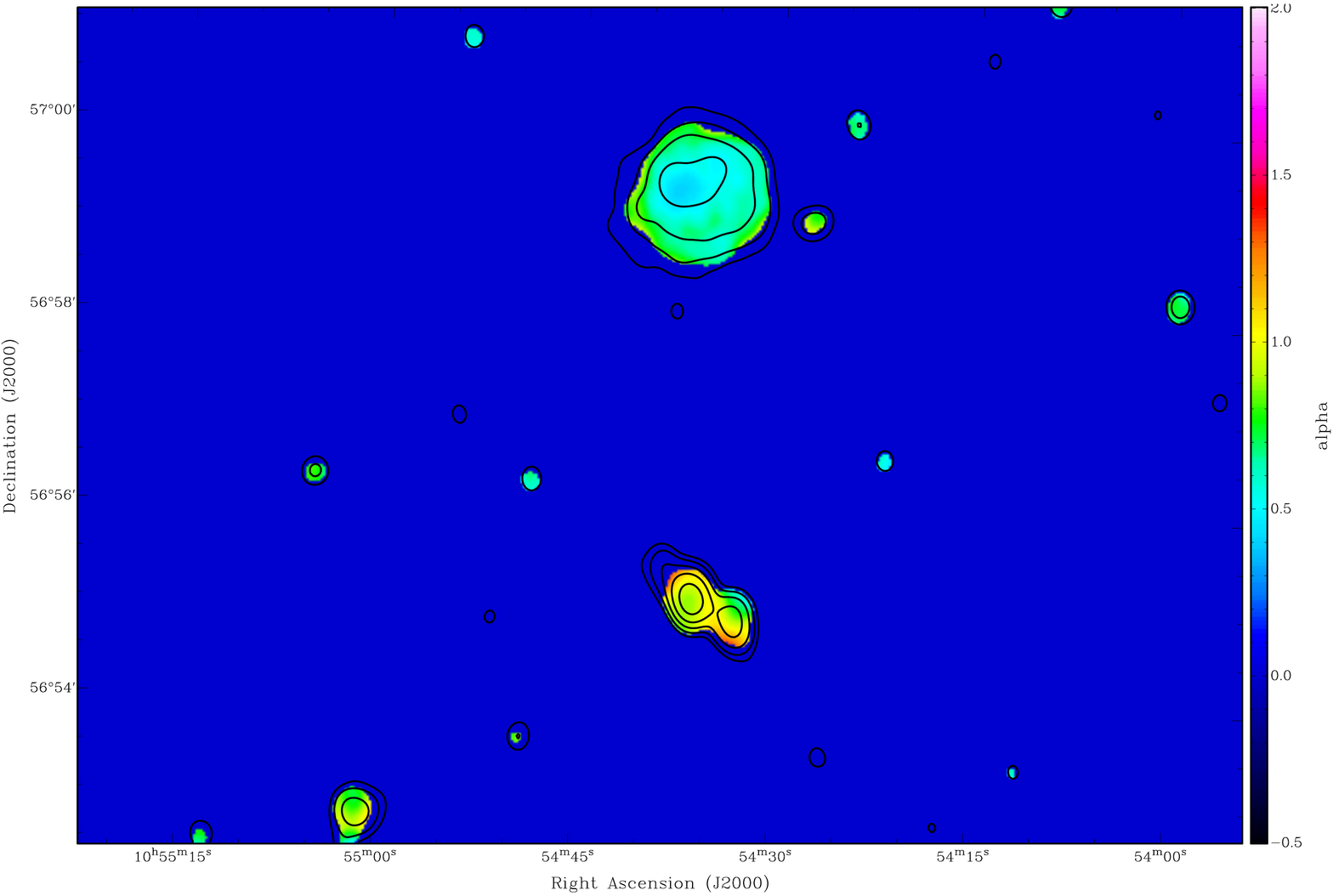}  
   \caption{Examples of images of the spectral index \si\ in various regions of the large Apertif-LOFAR image. The variety of spectral structures is evident. A few  double-lobed radio galaxies showing the classical spectral structure are seen, while the emission from a star forming galaxy and from a cluster object is also present.  Spectral indices commonly found in active radio galaxies (i.e.\ $0.5\lesssim \alpha \lesssim 1$, see e.g.\ \citealt{Gasperin18,Mahony16} and references therein) are indicated by the light green to yellow colours. Flat or inverted spectra ($\alpha \lesssim 0.5$) are indicated in blue and light blue. Extremely steep spectrum and USS values are indicated in red and pink.
 The contours represent the LOFAR emission, and contour levels are from $- 0.55, 0.55$ \mJybeam\ (5$\sigma$) increasing in steps of 2. Negative contours are shown in grey.}
\label{fig:doubles}
\end{figure*}

After implementing these corrections, a first spectral index image was produced by deriving the spectral index for all pixels with fluxes above 5$\sigma$ in both the 150~MHz LOFAR and the 1400~MHz Apertif images.
Examples  of the resulting spectral index images, showing the great diversity of the structure of the spectral index in various sources, are given in Fig.\ \ref{fig:doubles} and are discussed in Sect.\ \ref{sec:results}.

A map of the spectral index errors was also produced by using the formula 
\begin{equation}
\label{eq:error}
\sigma_\alpha = \frac{1}{\log (\nu_\mathrm{lofar}/\nu_\mathrm{apertif})} \sqrt{\left( \frac{\sigma_\mathrm{lofar}}{S_\mathrm{lofar}} \right)^2 +\, \left( \frac{\sigma_\mathrm{apertif}}{S_\mathrm{apertif}} \right)^2}
,\end{equation}
where $S_\mathrm{lofar}$ and $S_\mathrm{apertif}$ are the flux density pixel values on the LOFAR and the Apertif maps, respectively, and $\sigma_\mathrm{lofar}$ and $\sigma_\mathrm{apertif}$ are the noise levels of the respective data sets.  The 1$\sigma$ error on the detected spectral index ranges from values of $\sim 0.01 - 0.04$ for the brighter regions to $\sim 0.08 - 0.1$ for the fainter detected emission. 

In order to identify sources with extreme spectral properties, a second image was produced 
that contained lower limits to the spectral index for those pixels detected with  LOFAR (i.e.\ $S_\mathrm{lofar}\geq 0.55$ \mJybeam)  but which are undetected in the Apertif image at the 5$\sigma$ level ($S_\mathrm{Apertif} < 0.15$ \mJybeam). 
The lower limits to  the spectral index were estimated by using $S_\mathrm{Apertif} = 0.15$ \mJybeam\ in the spectral index formula: $\alpha = -\log (S_\mathrm{Apertif}/S_\mathrm{LOFAR})/\log (\nu_\mathrm{Apertif}/\nu_\mathrm{LOFAR})$. 
The two images -- spectral index detections and limits --  were combined into one image, which, for each object detected by LOFAR, shows the  spectral index for those pixels detected by Apertif as well as the limits where Apertif does not detect emission. 

\begin{figure*}
   \centering
 \includegraphics[width=7cm,angle=0]{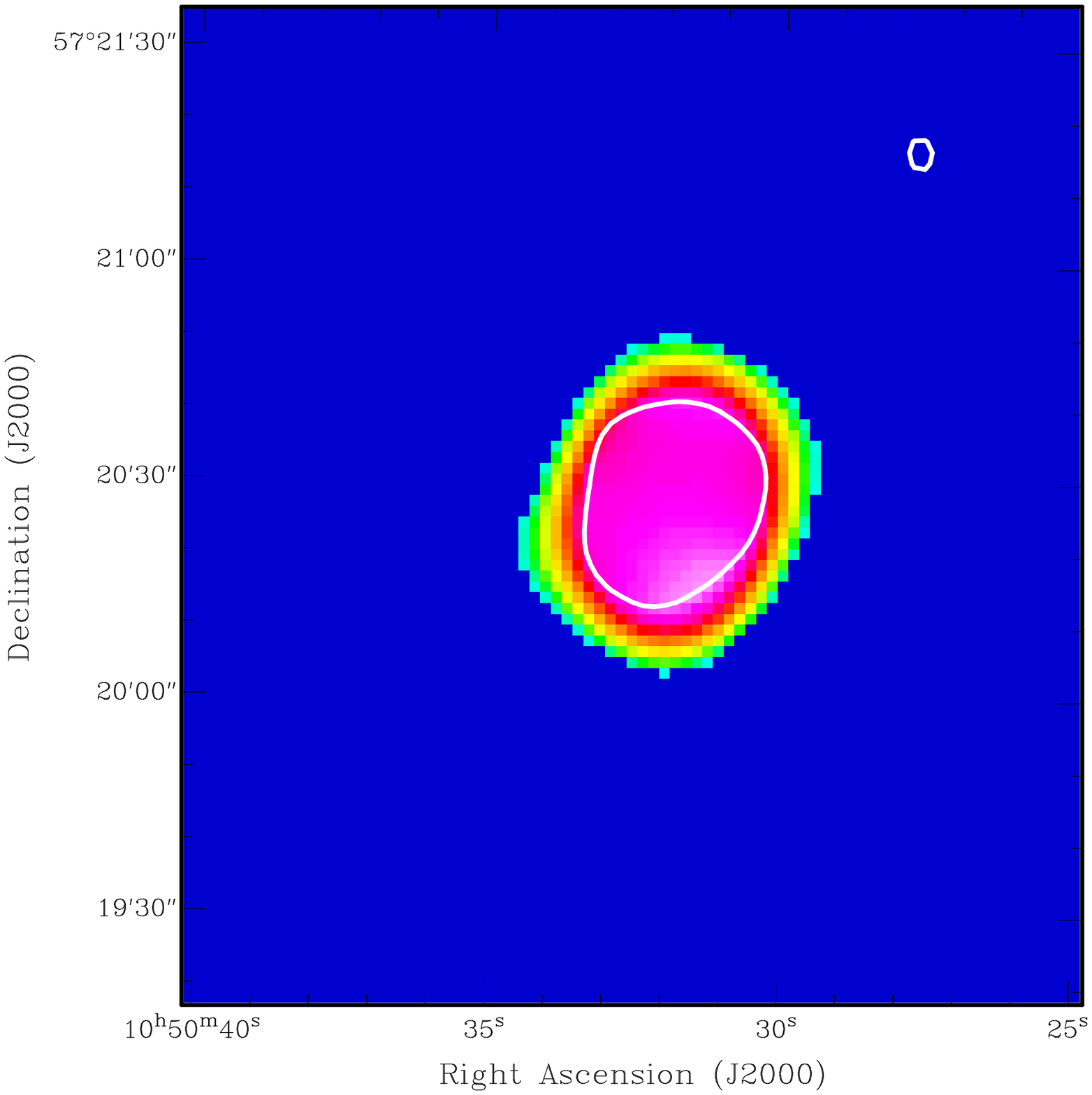} \hskip1cm
\includegraphics[width=7cm,angle=0]{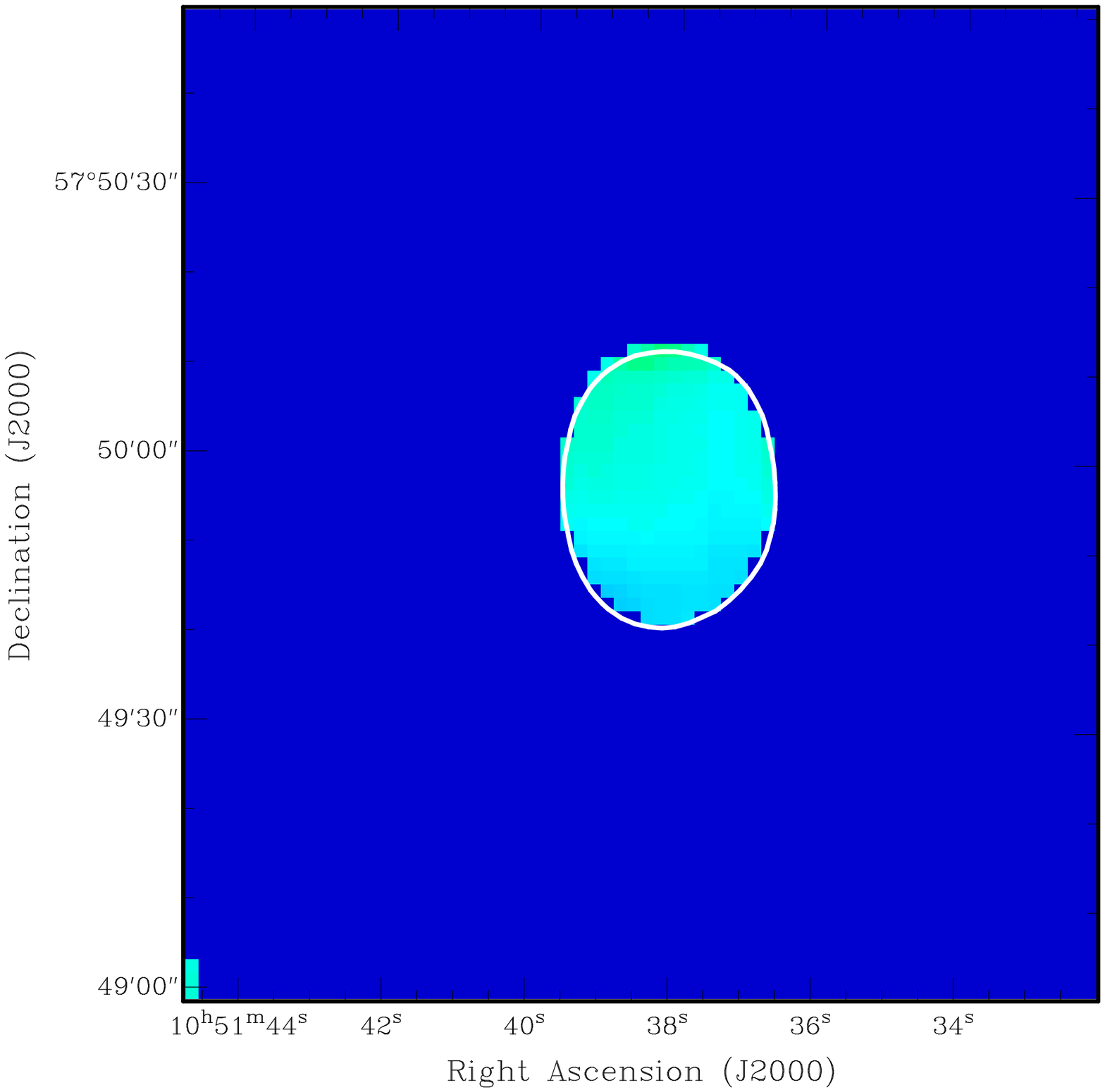} 

   \caption[]{Point sources in the combined spectral image (detection and lower limits) showing the effect of clipping of both the Apertif and LOFAR images at their respective  5$\sigma$ levels. The white contour shows the extent of the source in the Apertif image at the level of 5$\sigma$. Outside this contour, the spectral index represents a lower limit. The point source on the left has a spectral index of 0.9, causing the source to be brighter in terms of S/N in the LOFAR image, and, therefore, the region above 5$\sigma$ in the LOFAR image is larger than in the Apertif image. This creates an artificial region with a lower limit. The right-hand source has a spectral index of 0.54, which is close to the spectral index of 0.58 that corresponds to the noise levels in both images, so that the source has the same S/N in both images and no region with lower limits is created.}
\label{fig:point}
\end{figure*}

\begin{figure*}
   \centering
   \includegraphics[width=5.7cm,angle=-90]{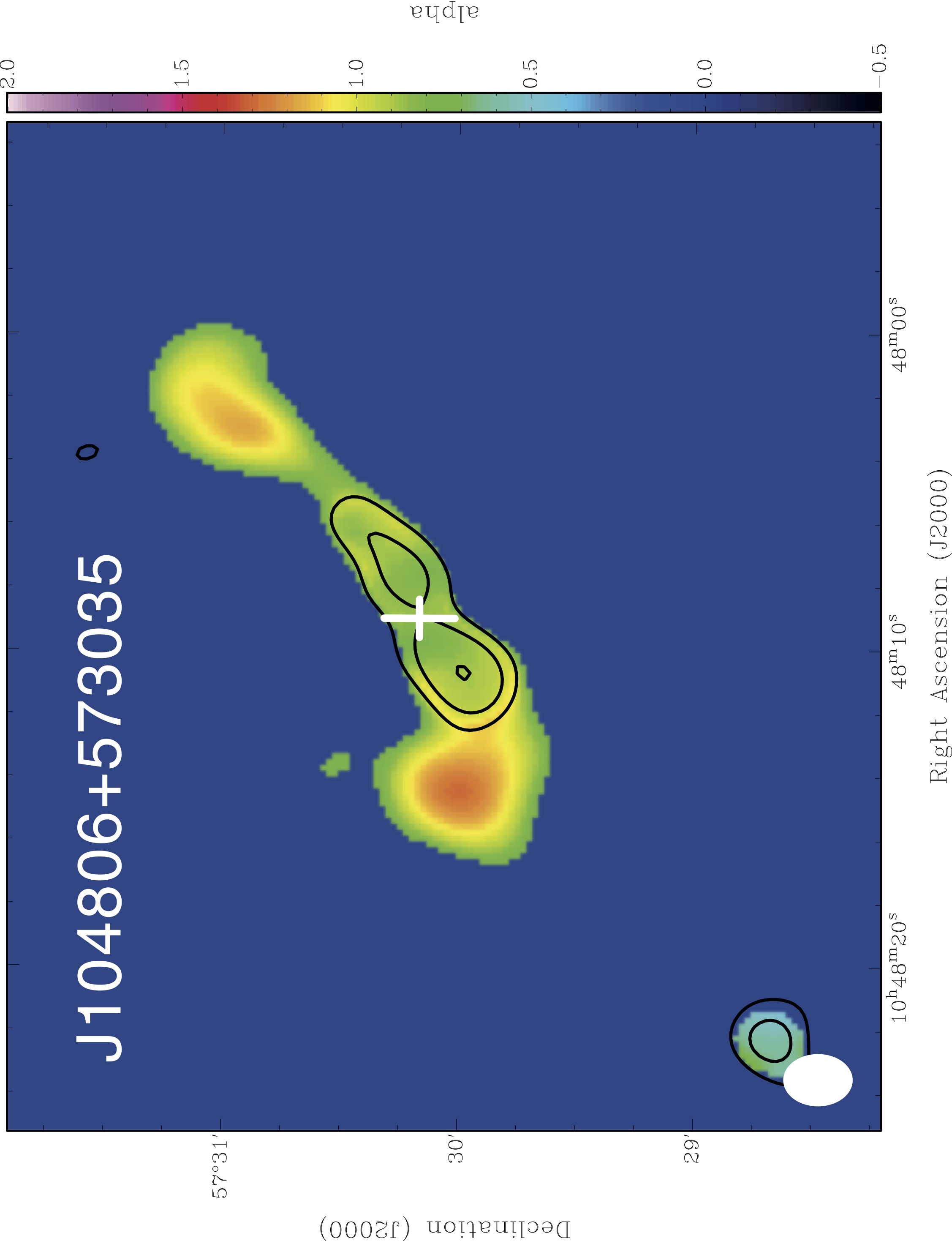}
   \includegraphics[width=5.7cm,angle=-90]{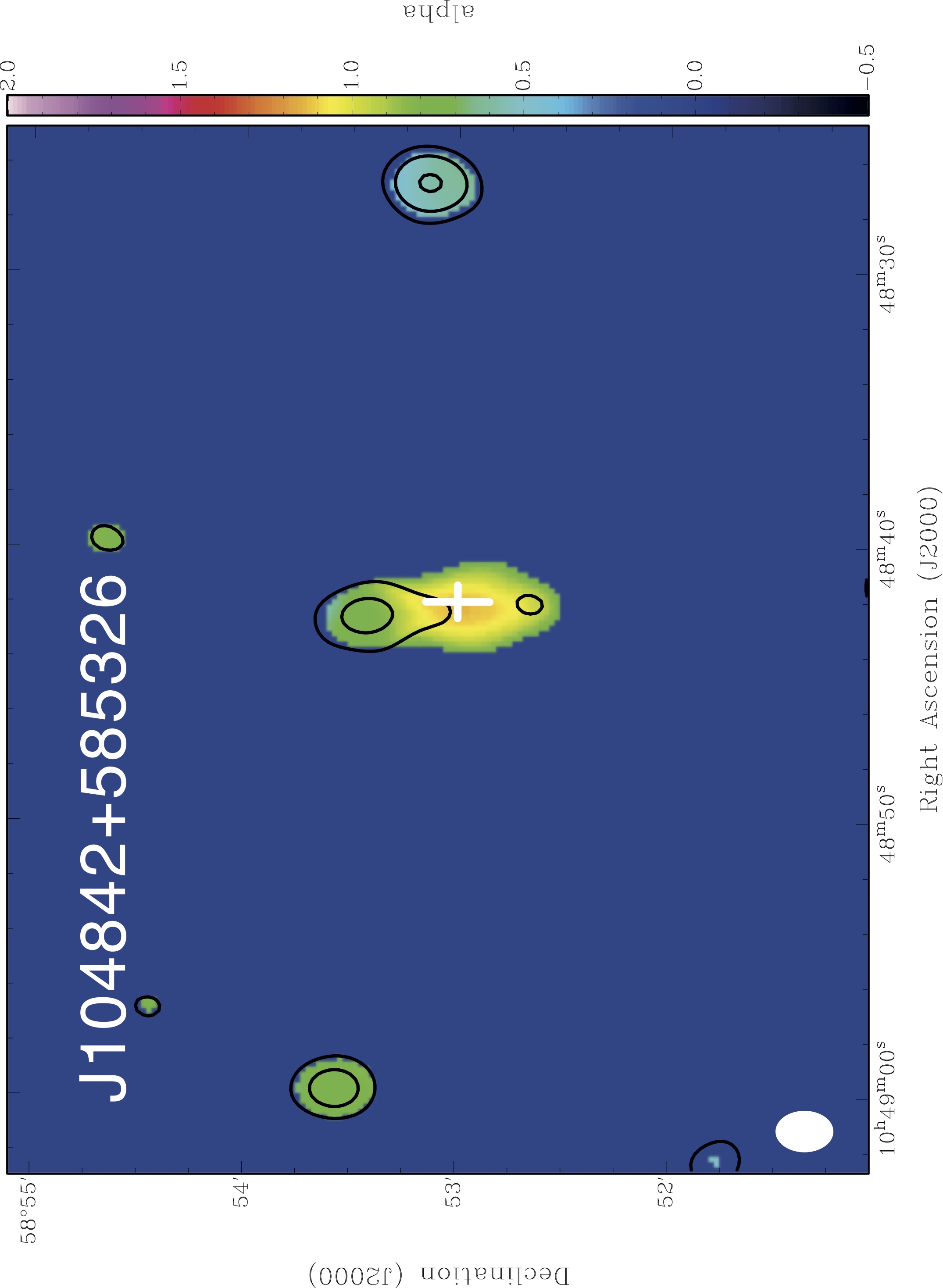} \\
   \includegraphics[width=5.7cm,angle=-90]{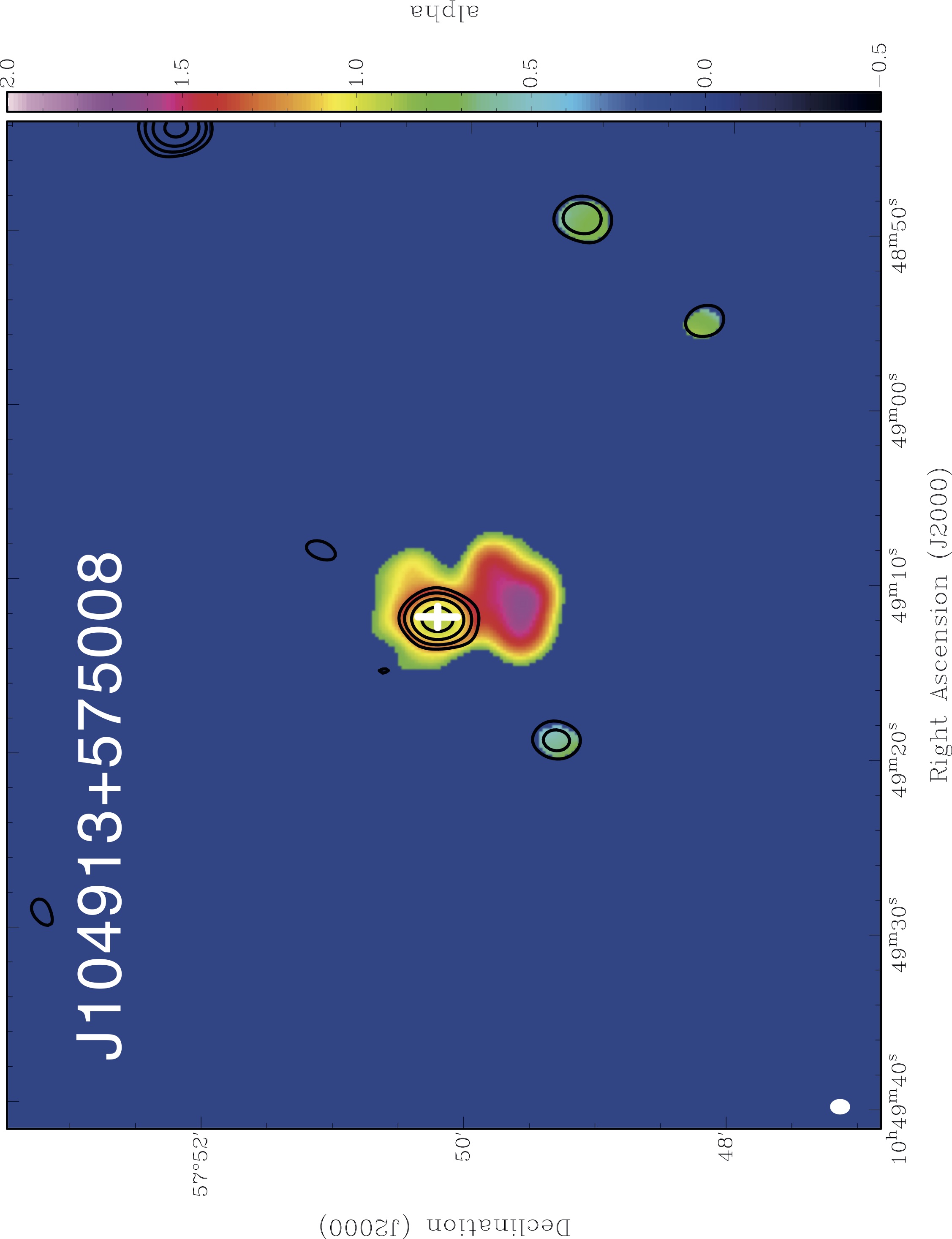}
   \includegraphics[width=5.7cm,angle=-90]{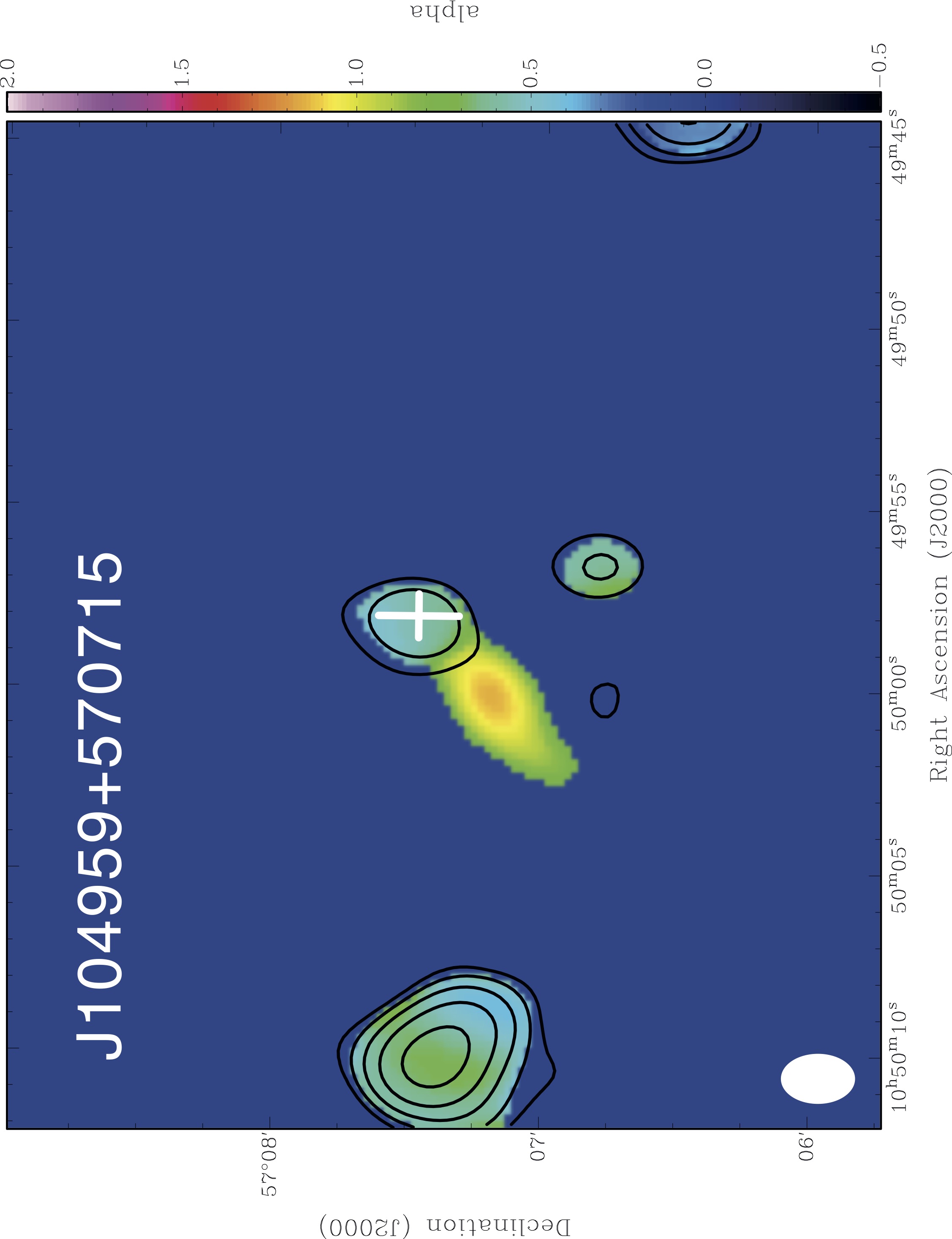} \\
   \includegraphics[width=5.7cm,angle=-90]{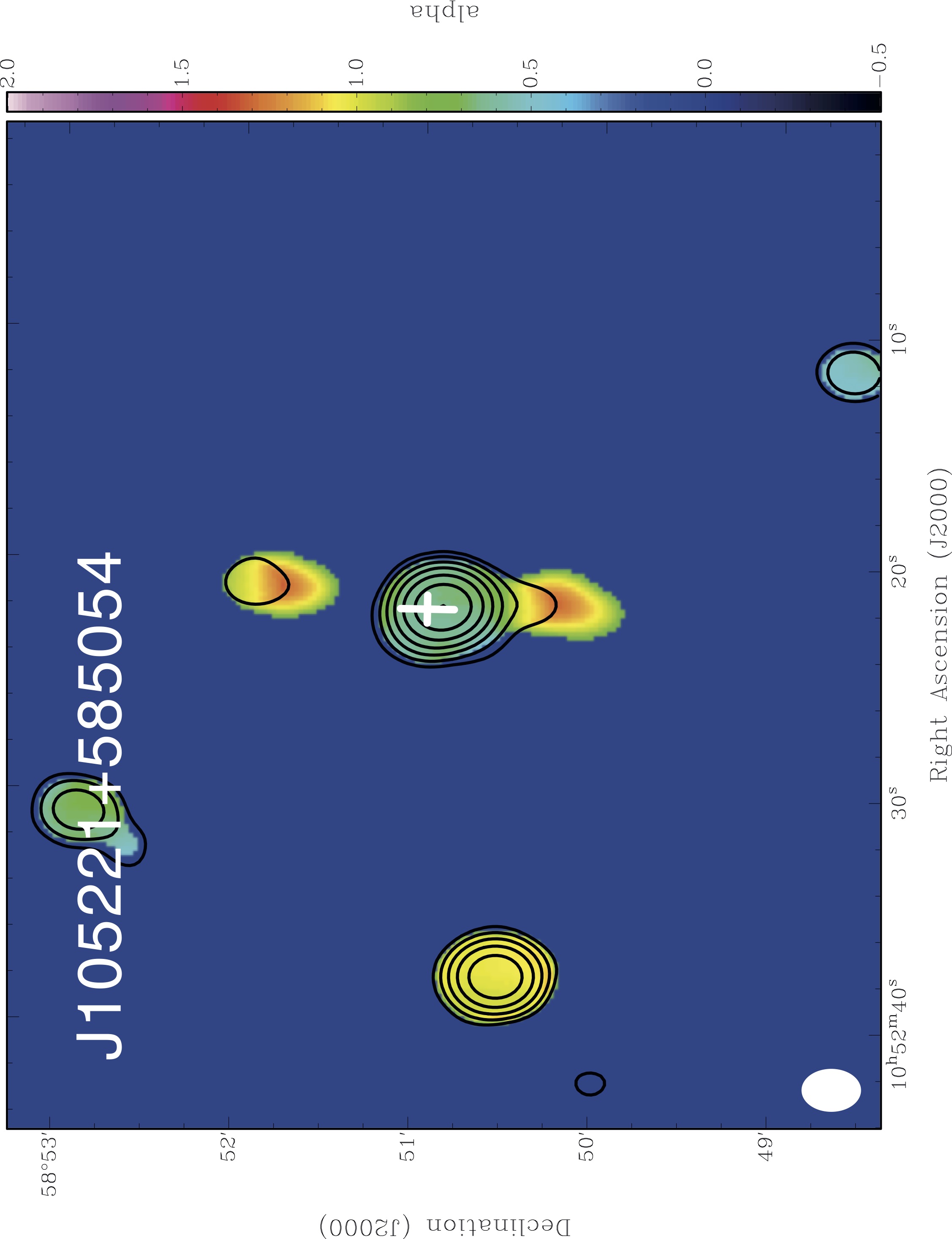}
   \includegraphics[width=5.7cm,angle=-90]{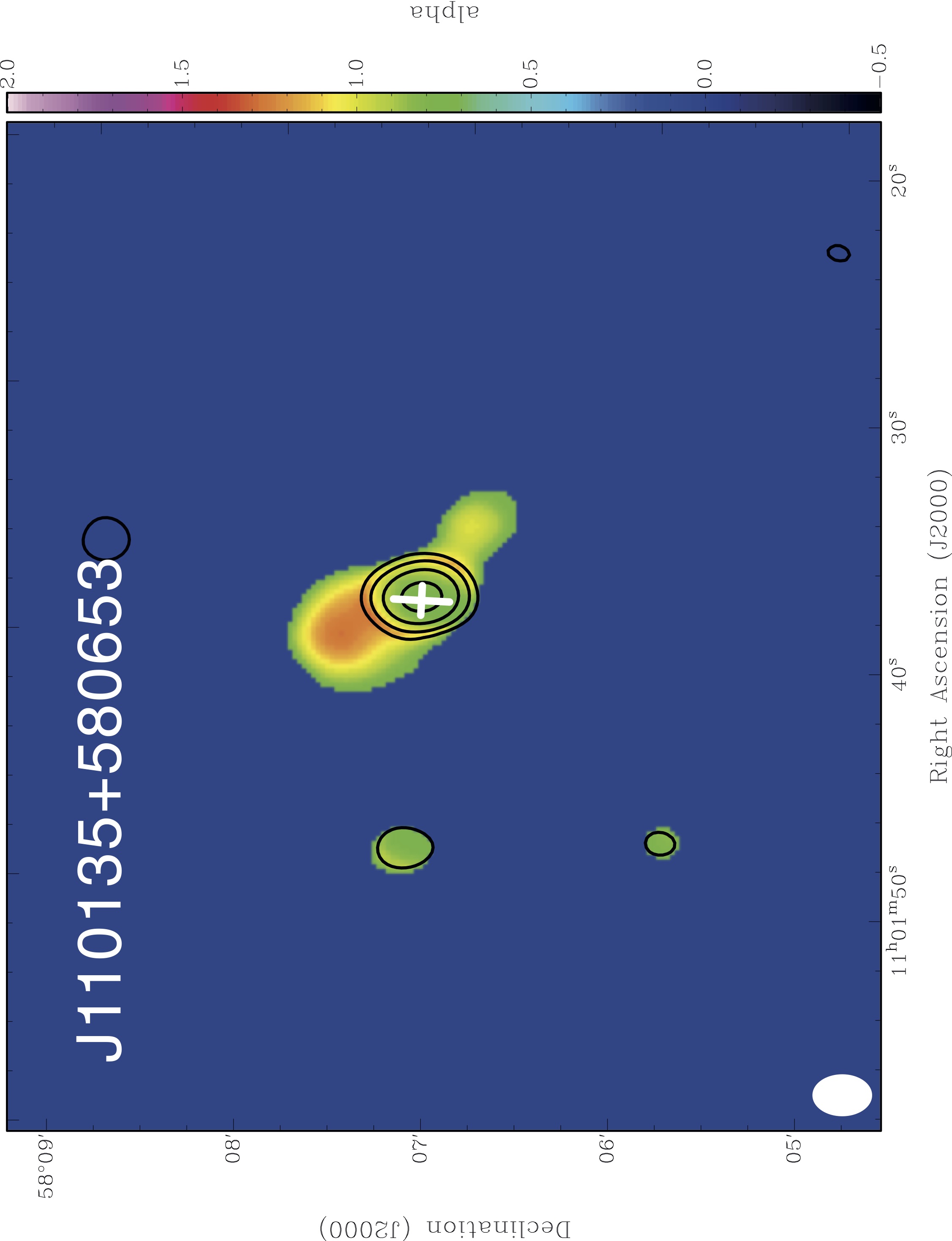} \\
   \includegraphics[width=5.7cm,angle=-90]{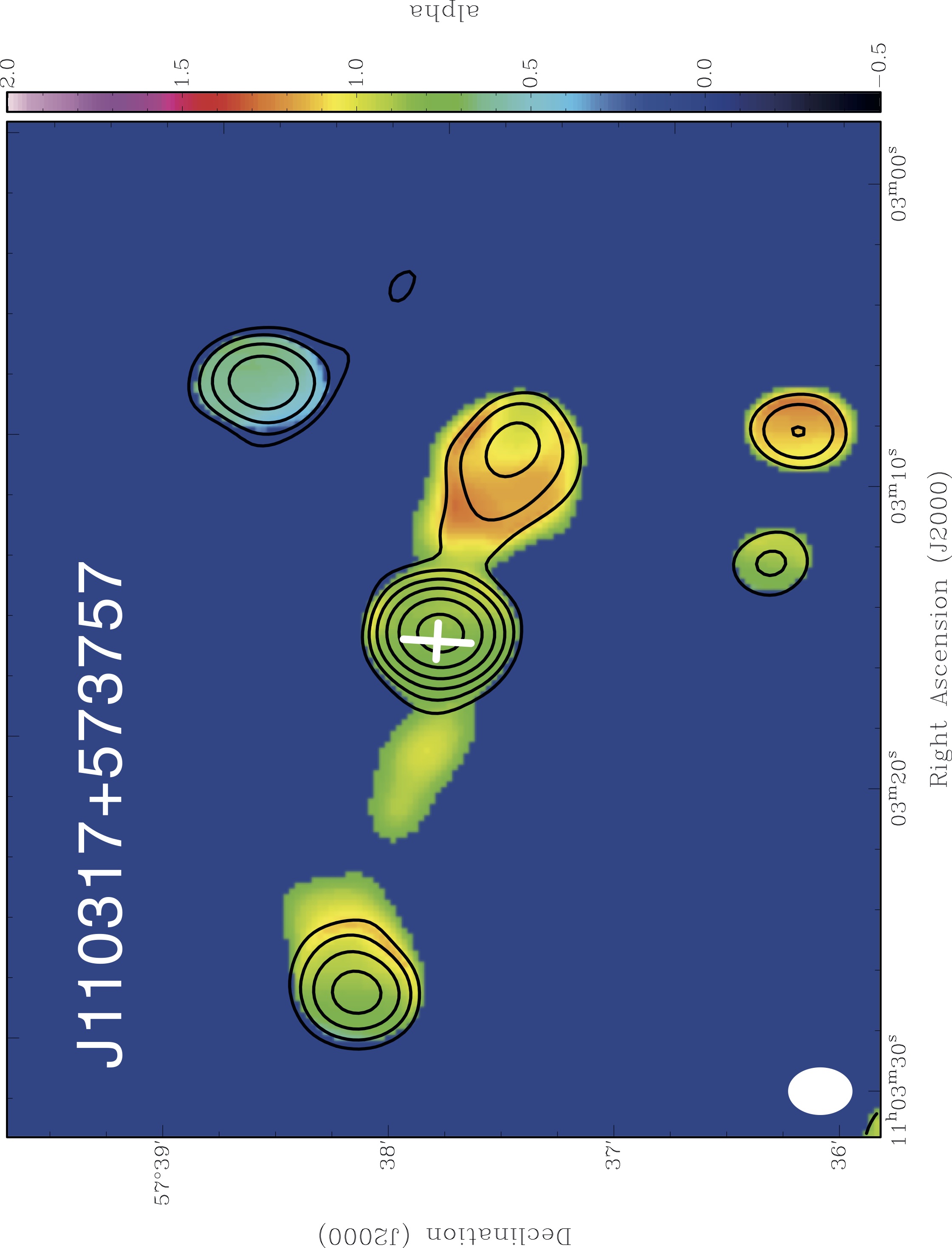} 
 \caption{Images of the spectral index and spectral index limits for the sources where part of the emission is USS. The contours represent the emission 5$\sigma$ at 1.4~GHz (starting at 0.15 \mJybeam\ and increasing with factors of 2). The spectral index inside those contours  are detections, while the ones outside are lower limits to the spectral index (due to the non-detection of Apertif; see the text for details).  The beam size is  $12^{\prime\prime} \times 16^{\prime\prime}$ and is indicated in the bottom-left corner. The edges around the sources  with a systematically flatter spectral index are the results of bias in the way the limits were derived. Further details are provided in Sect. \ref{sec:observations}.  The white crosses represent the optical identifications.
 }
\label{fig:partUSS}
\end{figure*}

As detailed below, the way we derived the limits on the spectral index  can introduce spurious features at the edges of sources (even in point sources). This can be seen for some of the sources (see Figs. \ref{fig:point}, \ref{fig:USS}, and \ref{fig:partUSS}) and should be kept in mind when interpreting these images.  For  regions where extended emission is detected in the LOFAR image, the flux density at 150~MHz is, as expected, decreasing towards the edges. Instead, for the undetected emission in Apertif, the upper limit to the 1400~MHz flux density used to compute the lower limit is kept fixed  at 5$\sigma$ (i.e.\ 0.15 \mJybeam). This combination gives an apparent flattening of the spectral index at the edges of the bright LOFAR emission, but  it just means that the lower limit to the spectral index is less stringent there because the LOFAR emission is fainter (Figs. \ref{fig:USS} and \ref{fig:partUSS}). This effect is particularly evident, for example, in the source J105233+585451. The mean spectral indices listed in Table \ref{tab:sources} are estimated taking the above effects into account and avoiding the affected regions. 

An additional effect is that for (point) sources that are detected by both instruments and have a spectral index steeper than 0.58 (i.e.\ the spectral index corresponding to the noise levels of the two instruments), the extent of the region above the noise clip in the LOFAR image is larger than that in the Apertif image. This is because the source is detected in the LOFAR image at higher S/N. This causes an artificial ring of lower limits around the point source, as illustrated in Fig.\ \ref{fig:point} and described in the caption. This  effect can be seen in some of the images.

\section{Results}
\label{sec:results}

A first, quick inspection of the spectral index image already shows that a large variety of structures is present. 
The entire field is too large to illustrate this in a single figure. Thus, in Fig.\ \ref{fig:doubles} we show  examples of interesting sub-regions where this variety can be appreciated. 
The images show the spectral index distribution of the detections for a number of different radio sources where both the 150~MHz and the 1400~MHz emission are detected above 5$\sigma$. 

A number of interesting objects can be seen. For example, a few cases of classical double, edge-brightened FRII objects appear prominently as the largest sources in the first four panels. Their spectral indices  follow the known trends: a flat-spectrum core (in the range $0 \lesssim \alpha \lesssim 0.4$), relatively flat spectrum jet head regions (including hot-spots as well as initial backflow), and lobes with its index steepening closer to the core, tracing the backflow of the radio plasma. 
These well-behaved structures demonstrate the reliability of the derived spectral indices and the good alignment of the images. Additionally, the values obtained are consistent with expectations (see e.g.\ \citealt{Harwood16} and references therein).
 
A number of flat or inverted spectrum sources are also visible in Fig. \ref{fig:doubles}. Most of them are unresolved and were already discussed by \cite{Mahony16}.
Two spiral galaxies are also seen, and a face-on one is visible in the bottom-right panel of Fig.\ \ref{fig:doubles} as the round object with a spectral index between 0.5 and 0.7. The bottom-left image includes a wide-angle tail, which is part of the Abell~1132 cluster (see also Sect.\  \ref{sec:A1132}). 

However, the objects most relevant for the present study are those characterised, partly or entirely, by  USS indices. 
As we described in Sect. \ref{sec:introduction}, 
USS indices are generally defined as \si\ = 1.2 or steeper (\citealt{Komissarov94}). Because we measure the low-frequency spectra, we relaxed this slightly in the case of regions where only limits on the spectral index are derived. Thus, we also consider sources with extended regions that have spectral indices steeper than \si\ $> 1$ as interesting USS candidates.

Following these criteria, the images were inspected by eye. We identified 15 sources dominated by USS emission.  For convenience, we separated these sources into two groups.  The first group includes sources where the entire emission is dominated by a USS.  The second group includes sources where the extended USS emission coexists with a region (mainly the central one) characterised by the typical spectral index of active radio sources (\si $ \sim$ 0.6 - 0.9). 
The selected sources are listed in Table \ref{tab:sources}, and they are discussed in more detail in Sects. \ref{sec:partlyUSS} and \ref{sec:USS}.
In Sect.\ \ref{sec:A1132}, we  briefly remark on the spectral indices derived for the region of Abell~1132. 

The optical identification for the selected sources was done  on the 6 arcsec LOFAR image, using the same procedure as described by \cite{Jurlin20} and Jurlin et al.\ (in preparation).  The redshifts were mostly obtained from the Sloan Digital Sky Survey (SDSS; spectroscopic and photometric) as described in \cite{Duncan19}. This catalogue  mainly covers  sources with $z<0.8$. 

Additional redshifts were determined using the new LoTSS multi-wavelength catalogue \citep{Kondapally20} and the refined hybrid photo-$z$ method  \citep{Duncan20}. These extra photometric redshifts, typically with $z > 1$ (marked with an asterisk in Table \ref{tab:sources}), are more uncertain as they lie outside the best multi-wavelength coverage area of the field.
For one object (J105554+563537), the photometric redshift is considered too uncertain, and we do not list it in Table  \ref{tab:sources}. 

\begin{table*}
\centering
\caption{List of extended sources selected based on their extreme spectral index properties (see Sect. \ref{sec:results} for details).  The spectral index properties are shown in Figs. \ref{fig:USS} and \ref{fig:partUSS}. Column 3 lists the spectral index at the location of the peak intensity oroptical position. Column 4 lists the average spectral index in the extended regions. These are derived by averaging over a box in the lobes, avoiding the region at the edges (see Sect. \ref{sec:observations}).}
\label{tab:sources}
\begin{tabular}{l|ccccc}
\hline\hline
 Source name  &   $z$ & \multicolumn{2}{c}{Average \si} & Log P$_{\rm 150~MHz}$ &  Refs \\
              &       & Central  & Lobes         & \WHz\ & \\
\hline
 \multicolumn{6}{c}{ Remnant radio sources\tablefootmark{a}} \\
 \hline
J104618+581421 & 0.168 & 1.3 & 1.1 - 1.5                   & 24.30 &   2     \\
J105034+564539 & 0.703 &  -  & 1.1 (N); 1.4 (S)             & 25.22 &        \\
J105123+582801 & $1.66^*$  &  -  & 1.2 - 1.5             & 26.50 &       \\
J105233+585451 & 0.468-0.505 & - &  1.5 to $>1.9$        & 25.17-25.22 &      \\  
J105554+563537 & -- &  -  & 1.8                   & 26.11 &  3     \\
J105723+565938 & 1.142 & 1.1 & 1.6 (N); 1.3 (S)              & 25.74 &  2     \\
J110201+583122 & 0.361-0.494 & - & $> 1.1$ (N);  $> 1.28$ (S)  & 25.50-25.74   & \\
J110255+585740 & 0.339 & -  & 1.2 (S); 1.6 to $>1.6$ (N)    & 25.35 &  3 \\
\hline
\multicolumn{6}{c}{ Restarted radio sources\tablefootmark{b}} \\
 \hline
J104806+573035 &  0.317 & 0.85  &  $>1.15$(N); $>1.3$(S) & 24.85    &  1,2  \\
J104842+585326 &  2.1685$^*$    &  --   & $>1.1$         & 25.53    &  2     \\
J104913+575008 &  0.073 & 0.90  & $>1.4$                 & 24.14    &  1,2  \\
J104959+570715 &  0.440 & 0.57  & $>1.1$                 & 24.59    &      \\
J105221+585054 &  0.978 & 0.64  & 1.0 to $>1.2$ (N); $>1.2$ (S) & 25.84  & \\
J110135+580653 &  0.1597 & 0.80 &  $>1.25$ (N); $>1.0$ (S) & 24.16    &      \\
J110317+573757 &  0.70  & 0.80  & 1.1 (W); $>1.0$ (E)      & 25.84    &      \\
 \hline\hline
\end{tabular}
\tablefoot{
\tablefoottext{a}{Extended sources with \si\ $> 1$ over the entire emission, see Sect. \ref{sec:USS} for details}; \tablefoottext{b}{Extended sources with partly USS emission, see Sect. \ref{sec:partlyUSS} for details.} 
$^*$ Redshifts from the expanded catalogue from Duncan  et al. in preparation, see Sect. \ref{sec:results}.
\\
{\bf References:} (1) \cite{Mahony16}; (2) \cite{Jurlin20}; (3) \cite{Brienza17}.}

\end{table*}

\subsection{Ultra-steep spectrum sources: Dying radio galaxy candidates}
\label{sec:USS}

\begin{figure*}
   \centering
    \includegraphics[width=5.7cm,angle=-90]{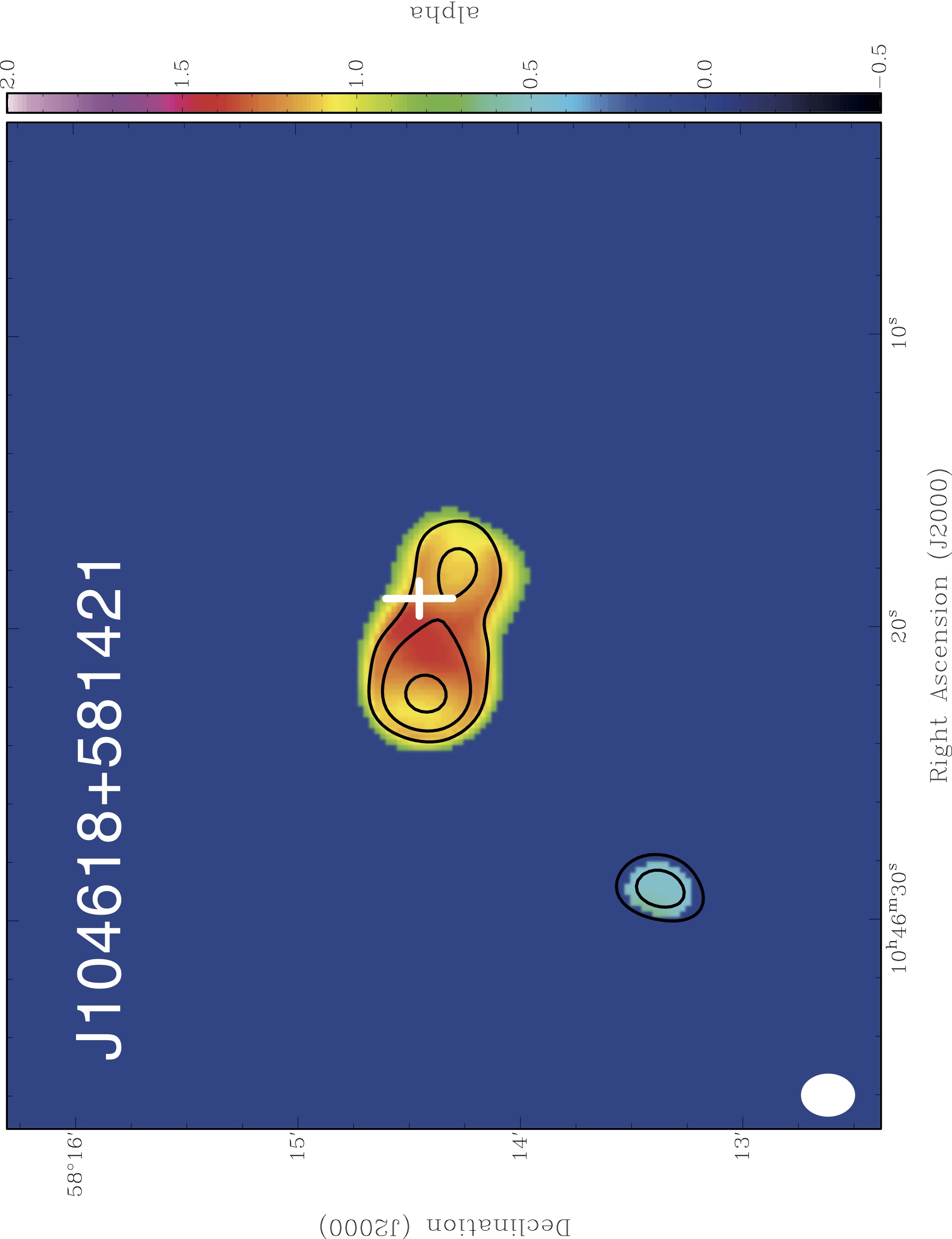}
   \includegraphics[width=5.7cm,angle=-90]{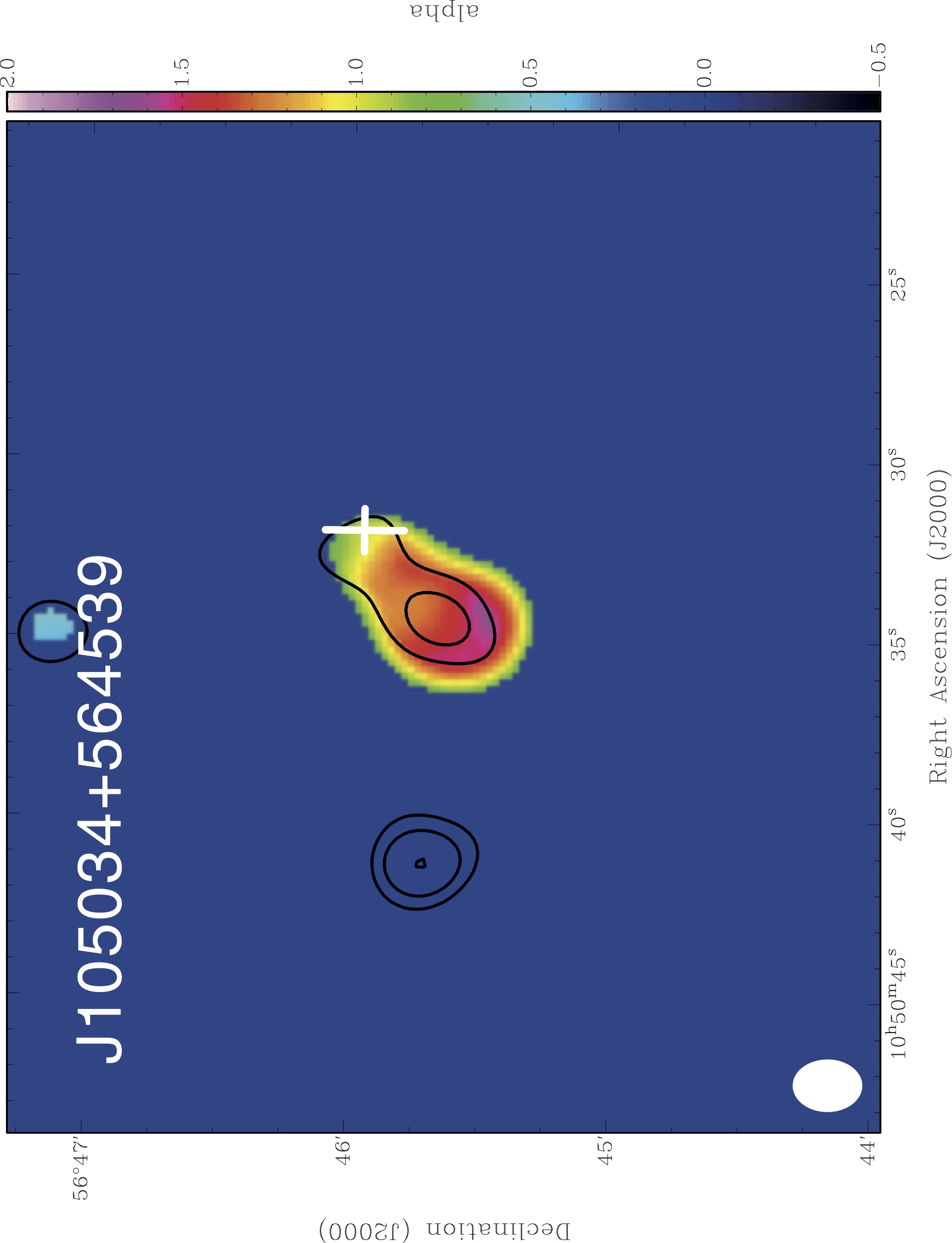}
 \includegraphics[width=5.7cm,angle=-90]{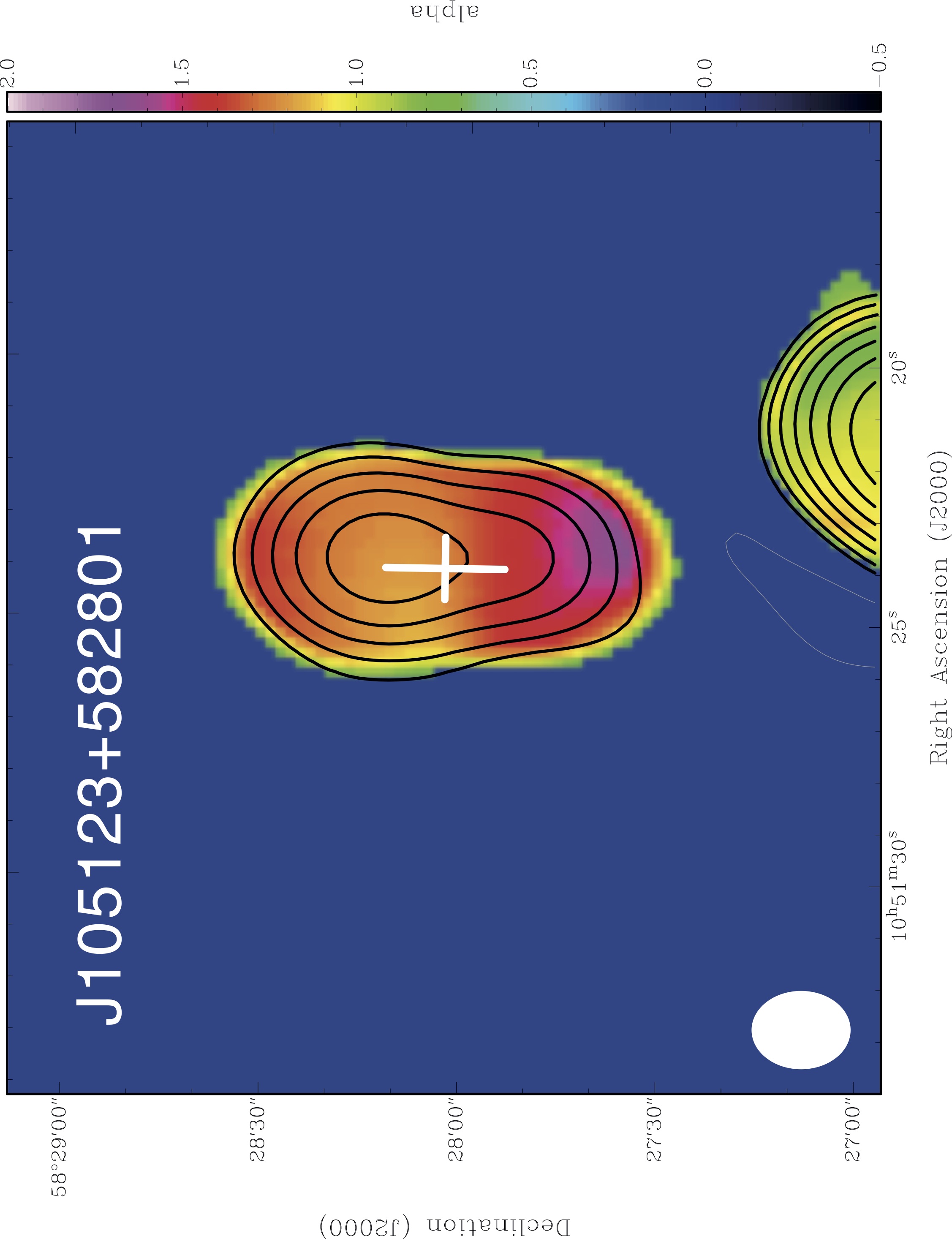}
 \includegraphics[width=5.7cm,angle=-90]{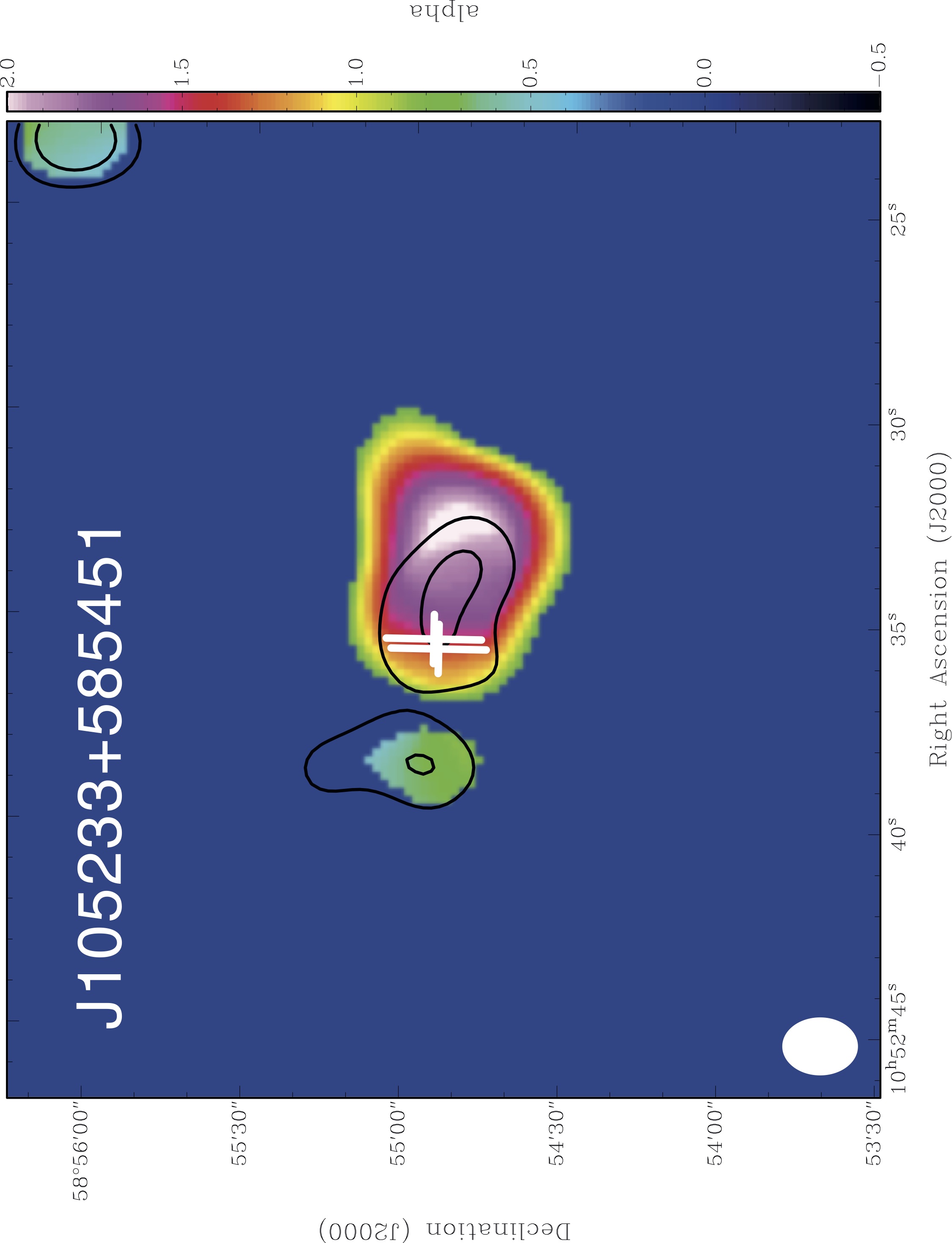}
 \includegraphics[width=5.7cm,angle=-90]{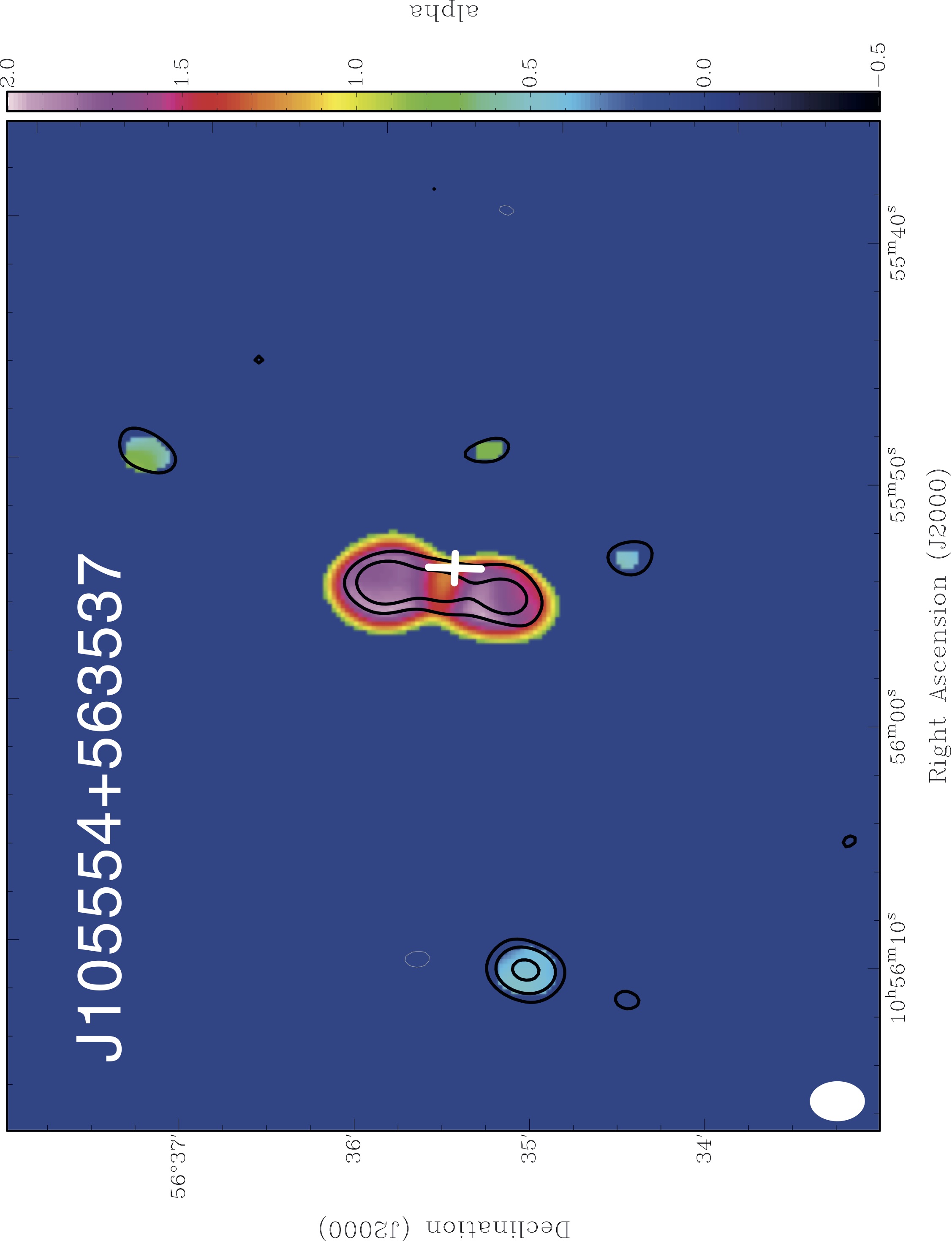}
 \includegraphics[width=5.7cm,angle=-90]{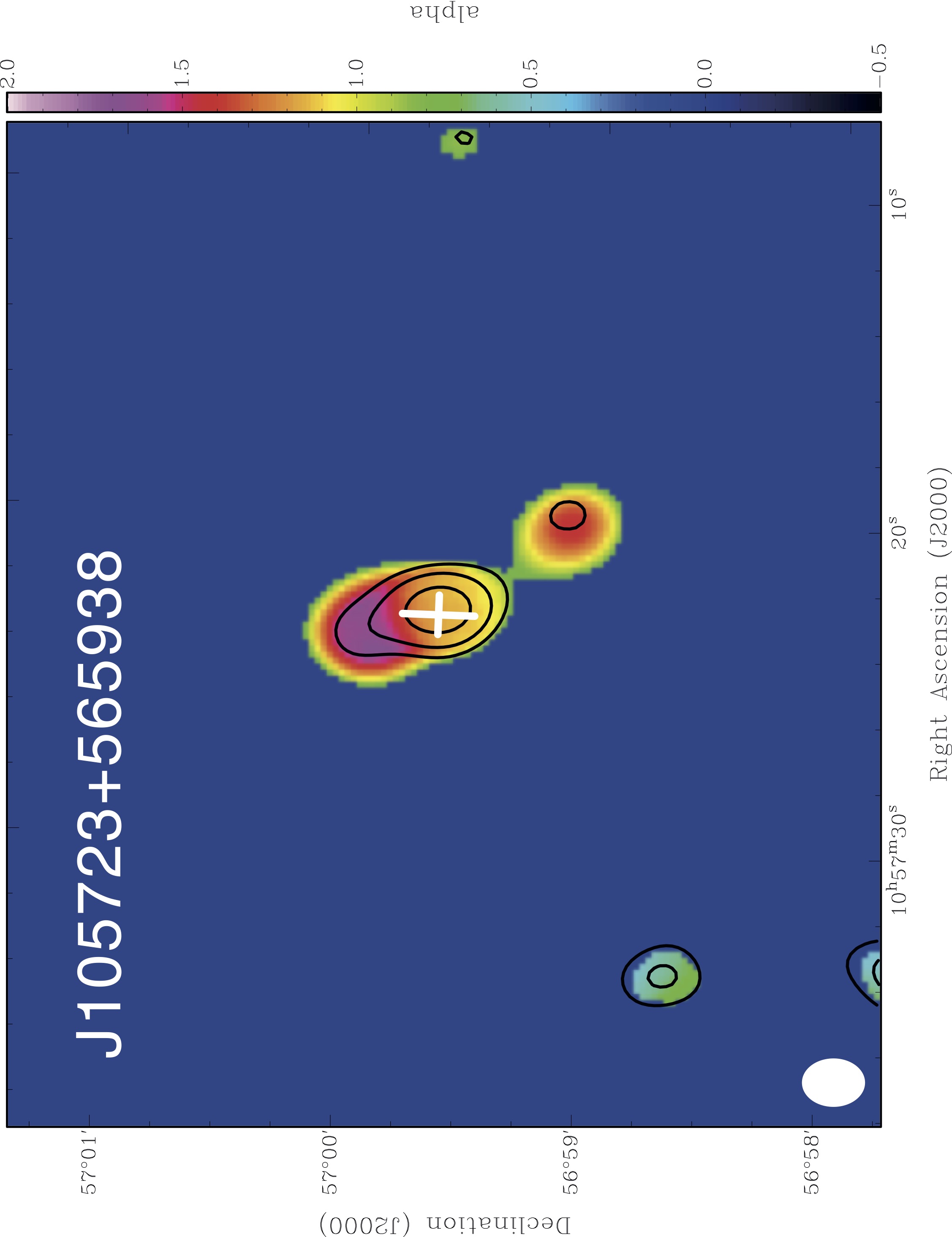}
 \includegraphics[width=5.7cm,angle=-90]{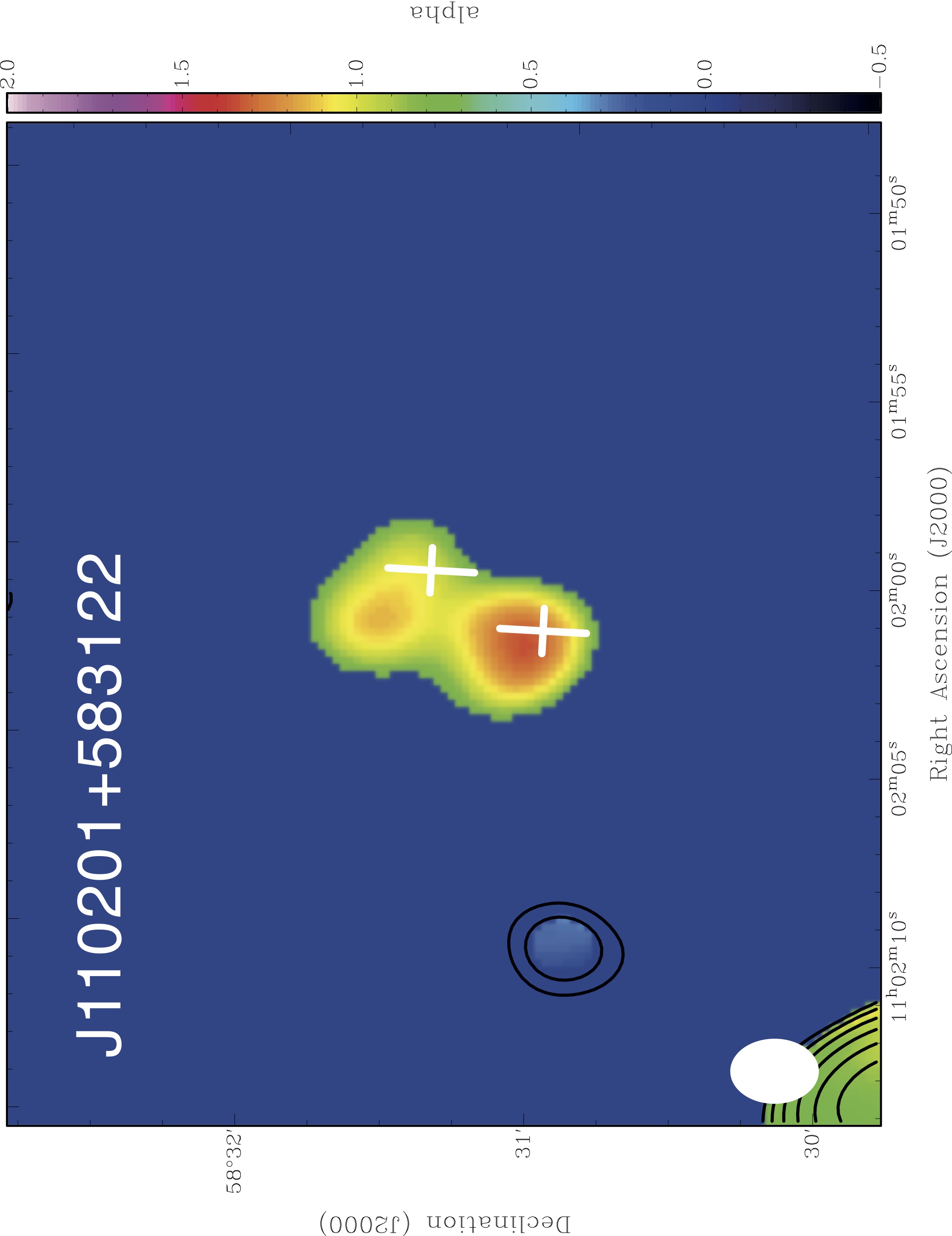}
 \includegraphics[width=5.7cm,angle=-90]{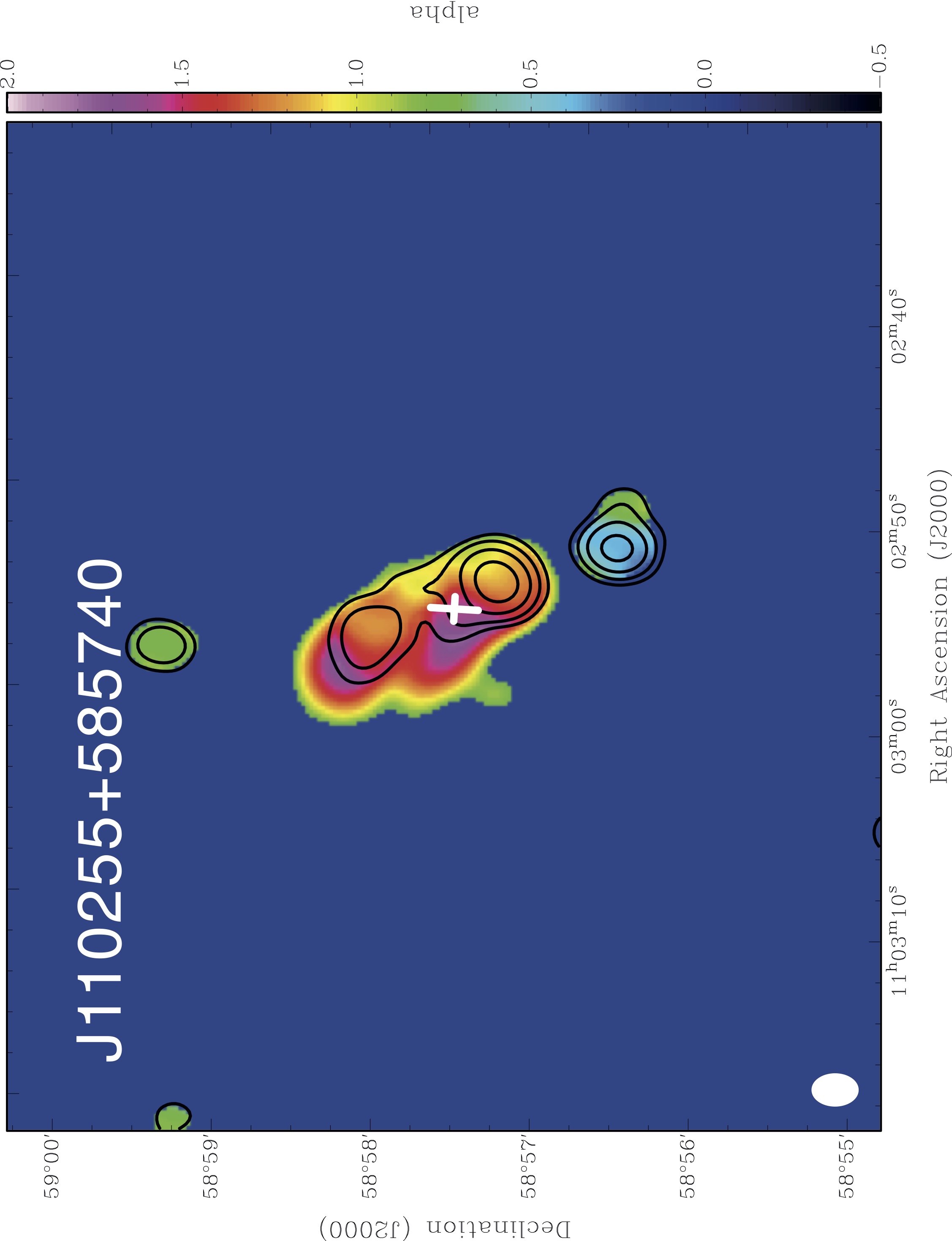}
   \caption[]{Images of the spectral index and spectral index limits for the sources characterised by USS emission. The contours represent the emission above 5$\sigma$ at 1.4~GHz (starting at 0.15 \mJybeam\ and increasing with factors of 2). The spectral index inside these contours  are detections, while the ones outside are lower limits to the spectral index (due to the non-detection of Apertif). See the text for details.   The beam size is  $12^{\prime\prime} \times 16^{\prime\prime}$ and is indicated in the bottom-left corner. The white crosses represent the optical identifications. In two cases, two potential identifications have been found (see also Table \ref{tab:sources}).  The edges around the sources of flatter spectral indices are the results of bias in the way the limits were derived. For more details, see the discussion in Sect. \ref{sec:observations}. }
\label{fig:USS}
\end{figure*}

We find eight sources where most, or all, of the emission is characterised by a USS.  Their spectral properties suggest that these are dying sources, where the extended and dominant USS is indicative of remnant structures left over from a past phase of the jet activity.  These sources are listed in Table \ref{tab:sources}, and their spectral index images are presented in Fig.\ \ref{fig:USS}. 

Three sources -- J105233+585451, J105723+565938, and J110201+583122 -- have the amorphous structure expected for remnant-like radio sources (see e.g.\ \citealt{Brienza17,Mahatma18}).   We note that 
J110201+583122 is not detected at all at 1.4~GHz; therefore, the structure of the spectral index presented in Fig. \ref{fig:USS} is defined by the LOFAR emission. Given the two candidate optical identifications (marked by the crosses in the figure), we also cannot completely rule out that this source is actually formed by two distinct radio AGN. 

Interestingly, a significant fraction of the extended USS candidate remnants are instead characterised by  double-lobed structures, namely\ J104618+581421,  J105554+563537,  J105723+565938, and J110255+585740.
One particularly interesting object is J105554+563537, which shows an extremely steep spectrum (\si$\sim 1.8$) but a double-lobed structure. This source (as well as J110255+585740) was already listed by \cite{Brienza17} based on the integrated spectral index.

The extreme steepness of the spectrum suggests an aged source that, nevertheless, appears to have maintained the double-lobed structure.  
Interestingly, the J105554+563537 radio galaxy has a core detected in deep VLA observations at 6~GHz (Jurlin et al.\ in preparation) and a candidate optical identification. The photometric redshift for this object is highly uncertain, but we cannot exclude a relatively high redshift\footnote{$z \sim 2$ based on available, but limited,  photometric data and requiring higher quality observations to be confirmed.}. 
High-$z$ radio galaxies are known to be often (albeit not always, see \citealt{Coppejans17}) characterised by USS emission \citep{Miley08}. Thus, in these objects, the steepening of the spectrum is often considered to be the result of a number of possible causes and not necessarily indicative of a remnant source. A number of suggestions have been made (see e.g.\ \citealt{Klamer06,Coppejans17} for overviews). One is that it is due to increased inverse Compton losses from scattering of cosmic microwave background (CMB) photons, an effect that is proportional to the CMB density and, therefore, to the redshift. Simulations of this effect have been presented by \cite{Morabito18} (see also \cite{Turner20}). While an impact from inverse Compton scattering cannot be excluded, the spectral index of J105554+563537 is extreme, even for high-$z$ radio galaxies. Therefore, we consider that the extreme spectrum of this object must be (at least partly) indicative of a remnant source. 

An important physical parameter that is usually extracted from the spectral information is the age of the source. Deriving reliable values requires  sampling  the radio spectrum at as many different frequencies as possible. This is clearly not the case for the sources presented here, with only two frequencies available. However, given that we are focusing on structures with extreme spectral index 'at low frequencies', we can derive limits to their ages by making some simple assumptions. 
To the first order, the ages  of the relativistic electrons ($t_s$) counting from the last injection (or re-acceleration) can be estimated -- neglecting expansion losses -- by employing the commonly used formula from \citep{Kardashev62}:

\begin{equation}
t_s=1590\frac{B^{\rm 0.5}}{(B^2+B_{\rm CMB}^2)\sqrt{\nu_{\rm b}(1+z)}} \\ {\rm Myr}.
\label{eq:time}
\end{equation}

\noindent This formula shows that  $t_s$  depends on the frequency of the break $\rm \nu_{\rm b}$ (in GHz), the strength of the magnetic field of the source $B$, and  the inverse Compton equivalent field $B_{\rm CMB}$, which depends on the redshift of the source as $B_{\rm CMB} = 3.25 \times (1 + z)^2$.

The magnetic field of the source is often derived based on the assumption of equipartition conditions between particles and the magnetic field. Under these conditions, the typical value of the magnetic field found in other remnant sources (i.e. $\sim  3 \rm \mu G$) can be assumed (e.g.\  \citealt{Tamhane15,Brienza16}). Some objects (such as FRII) are known to deviate from equipartition, and the magnetic field could be as low as $B \sim 0.3 B_{\rm eq}$ (see \citealt{Croston08,Ineson17,Turner18}). However, for such cases, the radiative cooling of the plasma is expected to be dominated by inverse Compton scattering of CMB photons, and  $B_{\rm CMB}$  dominates. Taking a range of redshifts between 0.15 and 1, where most of the sources are located, gives values of $B_{\rm CMB}$ between  4.3  and 13  $\rm \mu G$. 

For the  break frequency $\rm \nu_{\rm b}$, given that we look at structures with USS between 150 and 1400 MHz, the first assumption is that $\rm \nu_{\rm b}$ is below 1400~MHz. We can constrain this further by considering that, even assuming a steep injection index of the electrons (e.g.\ $\alpha_{\rm inj} = 0.8$), the  break frequency has to be at least as low as 600~MHz to achieve an \si \gta 1.0.  The break would be at lower frequencies (potentially even lower than 150~MHz) if the injection index wereflatter ($\alpha_{\rm inj} = 0.5$).

With a range in  break frequency between 150 and 600~MHz, and assuming $B_{\rm eq} = 3 \rm \mu G$, we derived ages up to between $\sim 160$ and $320$ Myr for the remnant structures. In most of the cases, the magnetic field will be dominated by $B_{\rm CMB}$ and the ages will become shorter (i.e.\ a few tens of megayears up to 100 Myrs), in particular for those with redshifts around or above 1.
The calculated values are consistent with simple considerations based on the lobe expansion speed. For old sources without strong bow-shocks, this will be of the order of the sound crossing time of the external gas, which is 100-200 Myr (depending on the environment) for a 200 kpc source.

Thus, these sources (or at least some of them) might represent the older tail of the age distribution of radio sources. This is in agreement with the results of the simulations by \cite{Shabala20} (see Sect.\ \ref{sec:relevance} for a discussion).


It is interesting to see, as mentioned above, that a number of the selected remnant sources have a double-lobed structure (i.e.\ J104618+581421, J105554+563537, J105723+565938, and J110255+585740).
This is relevant because the morphology of the selected remnant sources could give us some clues on their likely progenitors. Double-lobed sources  might be descended from FRII radio galaxies (or lobed FRI).
Indeed, their radio power, even at this evolved stage, tends to be higher than that of the typical FRI. 
These objects can be compared with the remnant B2~0924+35, for which \cite{Shulevski17}, using LOFAR, estimated an age of 50 Myr from the last re-acceleration and 120 Myr as the total age. In this object, while the structure appeared more amorphous than  the sources presented here, a flattening of the low-frequency spectral index is present at the extreme of the lobes, suggesting that these could have been the locations of the hot-spots when the source was active. Thus, an FRII was suggested as the progenitor for B2~0924+35 as well. For the objects presented here, high spatial resolution observations will be required to investigate whether this is the case for our sources. 

On the other hand, USS radio sources with a more amorphous structure (e.g.\  J105233+585451, J105723+565938, and J110201+583122) could represent the remnant phase of FRI. 
The possibility of an FRI progenitor has been proposed for, for example, the so-called Blob1, an amorphous remnant studied by \cite{Brienza16} that has also been confirmed as a remnant radio source thanks to a follow-up at high frequencies.  

In general, our observations confirm the variety of properties of the remnant objects. This will have to be taken into account when modelling the evolution of radio galaxies.

\subsection{Extended sources with partly USS emission: Restarted candidates }
\label{sec:partlyUSS}

\begin{figure*}
   \centering
   \includegraphics[width=15cm,angle=0]{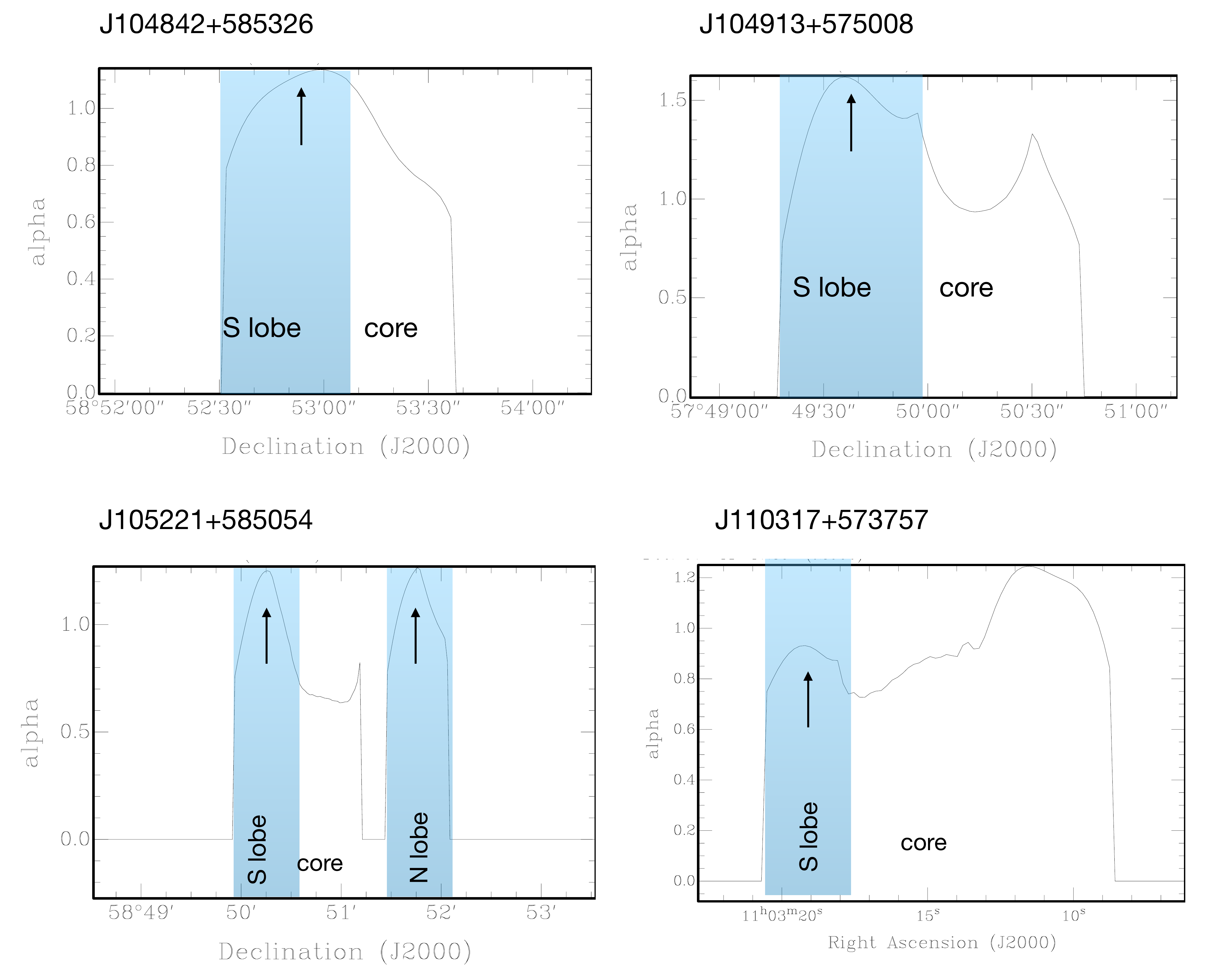}
\caption{Spectral index profiles across some of the sources presented in Fig. \ref{fig:partUSS}. The light blue areas correspond to regions of lower limits of the spectral indices. }
\label{fig:Profiles}
\end{figure*}

We have identified seven sources  that each show two regions with very different spectral index properties.  The first is a region (typically the central region) with a spectral index consistent with that of active galaxies.\ This first region is embedded in a second outer region that is characterised by a USS spectrum (or limit \si$>1.0$) that is typical of remnant structures, as discussed in the previous section. Thus, the spectral properties of these objects suggest that the external USS regions represent remnant structures that are left over from a previous phase of source activity, while the inner regions show the presence of a new, restarted phase of activity. 
The spectral index images of these selected objects are presented in Fig. \ref{fig:partUSS}. 
It is important to note that in many cases the spectral indices are actually lower limits. The spectral index distributions across the major axes of some of the sources are presented in Fig.\  \ref{fig:Profiles}. They clearly illustrate the values derived for the spectral indices and the region where they represent lower limits.  

We briefly describe the properties of every object. This is followed by some general considerations about the whole group.

{\it J104806+573035} is located in the area covered by the old WSRT mosaic and was already noted to have extreme spectral properties by \cite{Mahony16} and \cite{Jurlin20}. However, no analysis of the resolved spectral index was presented. This source has a structure made  by two double lobes, with the external ones being USS. To the first order, the structure resembles that of a double-double radio source, although it has the morphology typical of an FRI, with no hot-spots at the end of the lobes.  This, together with the USS outer lobes, are uncommon properties in double-double radio galaxies (\citealt{Mahatma19}). 

{\it J104842+585326} is a more uncertain case. It has very low surface brightness in the LOFAR image at 6 arcsec resolution and possibly has  a  double-lobed structure. Nevertheless, most of the structure shows  \si$> 1$.  Thus, based on the likely optical identification, the northern side (where the spectral index still has the typical value of 0.8) could actually be the location of a\ hot-spot. Therefore, it could possibly be an FRII source in the process of switching off instead of a restarted source.

{\it J104913+575008} is also located in the area covered by the old WSRT and was mentioned by \cite{Mahony16} and \cite{Jurlin20}. It is an intriguing source with a central, unresolved region surrounded by  highly asymmetric, USS diffuse emission and a lobe (undetected at 1400~MHz). It is hosted by a disk-like galaxy (see Fig.\ B.1 in \citealt{Mahony16}), which is quite uncommon  for extended radio sources. Even rarer, it has the spectral information of the extended emission.  The disk-like galaxy may suggest the presence of a gas-rich interstellar medium (ISM), where the jet (which in this case has relatively low power, $\sim 10^{24}$ \WHz)  may experience strong interaction and even (temporary) disruption. 
Thus, in this source,  the relatively bright core may also be the result of the interaction (e.g.\ similar to the case of PKS~1814-637; \citealt{Morganti11}). Another possibility is that this source is a switched-off classical double, with buoyancy directed towards the south. 
The real nature of this source will require follow-up observations (including of the environment).


{\it J104959+570715}  could represent a head-tail (HT) radio galaxy. If this is the case, it may belong to the galaxy cluster SWIRE CL J105004.0+570720  at a redshift of 0.43. 

{\it J105221+585054} is a double-lobed source with a prominent core, as can be seen in Fig. \ref{fig:partUSS_HR}. The lobes are USS and mostly undetected at 1400~MHz.  

{\it J110135+580653} is an asymmetric, double-lobed source. This object also appears to have a relatively prominent central region, as can be seen in Fig. \ref{fig:partUSS_HR}. Both lobes are undetected at 1400~MHz, and a spectral index limit of \si$> 1.1$ is found. 

{\it J110317+573757} is another   case of a source with a relatively prominent core, as can be seen from the higher spatial resolution LOFAR image shown in Fig. \ref{fig:partUSS_HR}.  The source shows two asymmetric lobes, the western of which has USS emission and a flatter spectral index (still at least \si $\sim 1$) at its edge, perhaps indicating the presence of a (remnant) hot-spot. The eastern lobe has a spectral index steeper than $\sim 1$. It is actually unclear whether the  eastern-most component is part of the source. 

\vspace{0.3cm}

In this group of sources, we were generally looking for cases similar to 3C~388 \citep{Roettiger94,Brienza20}, where the restarted activity can be identified only by the spectral properties and the sharp dichotomy in the spectral index across the source. Two regions with very different spectral index properties are observed; a sharp discontinuity in the spectral index distribution confirms the presence of two episodes of jet activity \citep{Brienza20}.  
Sources with these properties are considered rare, but our results show that they are not uncommon when appropriate sets of data are used.
In our sample, only J104806+573035 shows a clear transition to the USS indices occurring at a large distance ($\sim 100$ kpc) from the centre, similar to 3C~388. 
In the other sources, the new phase is still restricted to the region of the inner beam (corresponding to  $\sim 75$ kpc at the resolution of 14 arcsec for an average redshift $z\sim0.4$).


Four of the candidates in this group (J104913+575008, J105221+585054, J110135+580653, and J110317+573757) show relatively prominent nuclear or core emission.  This can be seen from their LOFAR images at full spatial resolution (6 arcsec) and presented in Fig. \ref{fig:partUSS_HR}.
This is interesting because \cite{Jurlin20} used this property (i.e.\ the core prominence) as a signature of restarted activity. Thus, this finding supports the use of this selection criterion (see \citealt{Jurlin20} for a full discussion).  

We can use the range of ages estimated for the remnant sources (i.e.\ between 160 and 320 Myr, see Sect. \ref{sec:USS}) as a reference for the outer regions of these restarted sources.  Because the central regions of these sources have spectral index characteristics of active sources, this means that, in this time interval (i.e.\ \lta 160 Myr),  the central SMBH has restarted. High-frequency observations of the restarted phase are necessary in order to estimate the actual length of the off phase. This has been done in the case of the radio galaxy 3C~388 (see \citealt{Brienza20} for details). For this object,
the derived total ages of the lobes is \lta 80 Myr, while the  period of inactivity in 3C~388 has lasted $\leq 20$ Myr. Obtaining such information for the objects presented here requires complementary high-frequency observations, which is the next (ongoing) step.  


\subsection{The region of Abell~1132}
\label{sec:A1132}

An analysis of the spectral index properties of Abell~1132 has been presented  by \cite{Wilber18}. LOFAR 150~MHz observations and GMRT  325 and 610 MHz data for this complex region were used to characterise the radio halo of the cluster and of the giant HT radio galaxy, both of which show USS.

With the Apertif image, we can extend this to 1400~MHz. The spectral index derived for this region is shown in Fig.\ \ref{fig:A1132} and confirms the USS properties of most of the structures in the region. For the outer part of the tail of the giant HT, only a limit of the spectral index could be derived, \gta 1.9, which is consistent with what is obtained at lower frequencies. The HT north of the cluster centre also has a very steep spectrum tail, \si $\sim 1.5$, that is detected by both Apertif and LOFAR. 


\section{Statistics of remnant and restarted sources selected by resolved spectral index}
\label{sec:statistics}

In addition to deriving their properties, obtaining  the actual  fraction of remnant and restarted radio sources is essential  for quantifying their life cycle. The LH area has recently been  used to start such a study. 
\citet{Jurlin20} created a sample  by selecting the radio sources with sizes larger than 60 arcsec. Mainly using the radio morphology, they identified candidate remnant and restarted radio sources from this sample and used the remaining sources as a comparison sample. In this way, they were able to derive the statistical  occurrence of remnants and restarted radio sources and, thanks to  detailed optical identifications, characterise their host galaxies (see \citealt{Jurlin20} for details). The results have also been compared with numerical simulations \citep{Shabala20}.

We can now further explore these results by adding the information presented here regarding the resolved properties of the radio spectra.  
Of the sources in the full sample selected by \cite{Jurlin20}, 43 are in the area covered by the Apertif image and, therefore, by our spectral study. From the 43, seven have been selected by our extreme spectral property criteria\footnote{For this analysis, we only consider sources  that fulfil the criteria applied by \cite{Brienza17} and \cite{Jurlin20}  to radio sources larger than 60 arcsec, based on the catalogue produced by the Python Blob Detector and Source Finder pyBDSF (\citealt{Mohan15}).}. They can be identified in Table \ref{tab:sources} from the references given in Column\ 7. 

Five had already been included in the sample of candidate remnant and restarted  radio sources presented by \cite{Brienza17} and \cite{Jurlin20}, either because of their morphology or because of available information about their spectral properties from the old WSRT mosaic, which, as mentioned above, covers   part of the area. 
Two (J104618+581421 and J104842+585326) were defined as active by \cite{Jurlin20}, but, according to our classification, they are a remnant and a restarted radio source, respectively. 
One source (J105723+565938), which was categorised as restarted by \cite{Jurlin20} based on its USS core,  is now found to have an entire USS structure  (see Fig.\ \ref{fig:USS});  therefore, we have classified it as a remnant. 

In summary, of the seven objects, three can be classified based on their spectral properties as restarted and four as remnant radio galaxies. 
Thus, although we were limited by the small number of objects, the spectral index information was crucial  for identifying or confirming the nature of all these objects as remnant or restarted radio sources.

 Using these objects, we can estimate the relative occurrence of remnant and restarted radio sources in our selected sample of extended sources. Based on the low-frequency spectral properties, we have, out of the 43 total sources, three restarted (7\%) and four remnant (9\%) sources. We can consider these fractions as lower limits on the occurrence of these sources because they were derived using only one criterion (i.e. low-frequency spectral index) for the classification (see below).  

Despite the uncertainties and low number statistics, our key result is that we confirm that a significant fraction of restarted and remnant radio sources are present in the sample of extended sources, as suggested by \cite{Jurlin20}. This also confirms the constraints provided by that sample to the models of \cite{Shabala20}. The results from the modelling (see below) are therefore further strengthened. 


Interestingly, four radio sources located in the area covered by this study were originally selected by \cite{Brienza17} as candidate remnant radio sources based on the low core prominence and low surface brightness, but they do not show extreme properties in their resolved spectral indices and they are not included in our selection presented in Table \ref{tab:sources}. 
These objects (J104646+56, J105230+56, J105703+58, and J110420+58) appear to have relatively normal spectral indices  at low frequencies. 
However, for one of them (J105230+56), a sharp steepening of the spectral index is observed at high frequencies, based on follow-up VLA 6~GHz observations that will be presented in Jurlin et al.\ (in preparation). Thus,  this object is a genuine remnant that does not show USS at low frequencies (yet).  This likely represents a younger remnant that only recently switched off (as discussed in \citealt{Godfrey17,Brienza17,Hardcastle18,Shabala20}) or, perhaps, is the result of a different combination of  on and off times, as was found for Blob1 \citep{Brienza16}. 
This finding is consistent with  recent results showing that the spectral indices of remnants are not always extremely steep at low frequencies, as is usually assumed. 
The variety of their properties shows that a selection based only on the low-frequency spectral index will not select all remnant and restarted radio galaxies, though it does select most of them (as shown by \citealt{Hardcastle18}). This also indicates the need for a variety of diagnostics to identify all of them. 

Unfortunately, no high-frequency deep observations are available for the other three objects. 
Their final classification will therefore remain uncertain.

In summary, although limited to a small number of sources, this analysis shows the possibilities offered by resolved spectral information for selecting remnant and restarted radio sources. When applied to the data that will be obtained over the large areas covered by the full Apertif and LOFAR surveys, it will provide a wealth of information on the different evolutionary stages of radio sources.

\section{Relevance for the life cycle of radio galaxies}
\label{sec:relevance}
 

As discussed in \cite{Jurlin20}, the comparable fraction of remnant and restarted radio sources suggests that the restarted phase can, in some cases, follow a relatively short remnant phase.  This is also suggested by some single-object studies available in the literature  (e.g.\ \citealt{Parma10,Murgia11,Saripalli12,Konar13,Bruni20,Brienza20,Maccagni20}). These studies find  lengths of the inactive period between two phases of activity of between a few megayears to a few tens of megayears. 
However, the fact that in the present sample we do not see a restarted phase in the majority of the remnants is telling us that the restarted phase does not always follow shortly thereafter. Thus, a variety of situations and timescales can be found for the restarted phase as well.

Our results are particularly relevant for constraining the predictions from  modelling radio source evolution. The models presented in \citet{Brienza17}, \citet{Godfrey17}, and \citet{Hardcastle18} suggest that the observed fraction of remnant sources can be explained if these structures fade relatively fast (on timescales of at most a few $ \times 10^8$ yr) due to  both radiative losses and dynamical expansion. 
This has recently been  confirmed by more sophisticated modelling presented in \cite{Shabala20}. They employ forward modelling with the Radio AGN in Semi-analytic Environments code (RAiSE, \citealt{Turner15})  to constrain the distributions of the lifetimes and  kinetic jet powers of  active radio galaxies.  The models are used to  make predictions for the size of the remnant and restarted populations, and these are compared  with the observations of \cite{Jurlin20}.  Two types of models are considered: constant age models (i.e.\ models where all sources live to the same age) and models with  a power-law distribution for the ages ($p(t_{\rm on}) \propto t^{-1}$). 
 The former models did not provide a good description of the number of candidate remnant and restarted radio sources observed by \cite{Jurlin20}, although they were still  marginally consistent with them. The best fit was, instead, obtained with  power-law models  with a high fraction of short-lived sources (i.e.\ ${t_{\rm on} < 100}$ Myr) in order to reproduce the observed fraction of remnant and restarted radio sources (see \citealt{Shabala20} for details). 
 
 The results presented here not only put stronger constraints on the models and strengthen the conclusions of \cite{Shabala20}, but also suggest that we are selecting the remnant structures that are in the older tail of the distribution. This is in contrast to the remnants selected using morphological properties (see \citealt{Brienza17}), which are likely observed soon after the switch-off of the radio source and are expected to evolve quickly due to dynamic expansion. The sources we selected  have the entire structure (many tens to hundreds of kiloparsecs in size) characterised by USS emission: The switch-off of the activity has already affected the entire source. This is also in agreement with  the fact that  a restarted phase is already seen in some of the sources.


\section{Conclusions and future outlook}
\label{sec:conclusions}

We have presented a pilot study of the resolved spectral index derived between 150 and 1400~MHz for  radio sources in an area of the LH region of about 6 deg$^2$. Exploring the spectral index at low frequencies has allowed us to identify   15 sources where all (or most) of the extended structure is dominated by USS. Such steep, low-frequency spectra  are clear signatures of very old plasma where there is no ongoing replenishment of fresh electrons. Therefore, they are key to providing constraints on the age and life cycle of the radio activity. This study represents a step forwards in understanding this cycle, complementing studies of radio galaxies in  the extreme phase of their life, which are selected using  morphology. The complex picture we are trying to put together shows that a variety of diagnostics is needed to identify radio sources in various stages of their evolution. 

The results presented here  were obtained by using just one pointing of the Apertif surveys, which had only 6 deg$^2$ in common with  data from the ongoing LOFAR surveys that were available at the time of writing. This highlights the possibilities that are opening up as the Apertif survey builds up to the final survey area of 3000 deg$^2$ (Adams et al.\ in preparation), all of which will be enclosed by the area covered by the LOFAR surveys. Our study shows that this combination with the data from the 150-MHz LOFAR survey (LoTSS, \citealt{Shimwell19}) will give  valuable new information on the spectral properties of radio sources, something that will be even more the case when the data from the  upcoming LOFAR low-band survey at 54 MHz (de Gasperin et al.\ in preparation), with resolution well-matched to Apertif, become available. This will make it possible to expand this kind of study to much larger samples covering a large variety of types of radio sources, including rare ones.  Building such large samples will allow us to connect the observed properties of the life cycle to the properties of the host galaxy as predicted  by models.

Handling the large samples that will be obtained with the combined Apertif and LOFAR surveys will require developing automatic tools for the  characterisation and classification of the spectral properties of radio sources and for the selection of the sources of interest. Such tools can also be connected to automatic ways of exploring the morphological classification of radio sources, which are now starting to be developed using automatic techniques (see e.g.\ \citealt{Mingo19,Ralph19,Galvin19}, Mostert et al.\ submitted).

As a final remark, it is worth noting that the full understanding of the life cycle of radio sources will also require connecting other properties of the host galaxy, such as the presence and  the properties of the gas. For low-redshift objects, information on the presence of \HI\ can be derived from the same Apertif observations as these also provide data at high spectral resolution.   If the  higher fraction of \HI\ detection in restarted radio galaxies, as suggested by studies of single objects (e.g.\ \citealt{Saikia09,Morganti18}), is confirmed for larger samples, it would suggest a link between cold gas and restarting nuclear activity. 
A wider range in redshift can be  probed for the ionised gas, either using SDSS data or upcoming WHT Enhanced Area Velocity Explorer (WEAVE) observations. The connection between the evolutionary stage of the radio source and the reservoir of gas will undoubtedly bring new and interesting insights.

\begin{acknowledgements}
This work makes use of data from the Apertif system installed at the Westerbork Synthesis Radio Telescope owned by ASTRON. ASTRON, the Netherlands Institute for Radio Astronomy, is an institute of the Dutch Science Organisation (De Nederlandse Organisatie voor Wetenschappelijk Onderzoek, NWO).
LOFAR, the Low Frequency Array designed and constructed by ASTRON, has facilities in several countries, that are owned by various parties (each with their own funding sources), and that are collectively operated by the International LOFAR Telescope (ILT) foundation under a joint scientific policy. 
The research leading to these results has received funding from the European Research Council under the European Union's Seventh Framework Programme (FP/2007-2013)/ERC Advanced Grant RADIOLIFE-320745. JMvdH acknowledges funding from the Europeaní Research Council under the European Union’s Seventh Framework Programme (FP/2007-2013)/ERC Grant Agreement No. 291531.
IP acknowledges support from INAF under the SKA/CTA PRIN FORECaST and the PRIN MAIN STREAM SAuROS projects.
MB acknowledges support from the ERC-Stg DRANOEL, no 714245. LCO acknowledges funding from the European Research Council under the European Union's Seventh Framework Programme (FP/2007-2013)/ERC Grant Agreement No. 617199. EAKA is supported by the WISE research programme, which is financed by the Netherlands Organization for Scientific Research (NWO). JvL acknowledges funding from the European Research Council under the European Union's Seventh Framework Programme (FP/2007-2013) / ERC Grant Agreement n.\ 617199 (ALERT), and from Vici research programme ARGO with project number 639.043.815, financed by the Dutch Research Council (NWO).
\end{acknowledgements}

\bibliographystyle{aa}

\begin{thebibliography}{17}
\expandafter\ifx\csname natexlab\endcsname\relax\def\natexlab#1{#1}\fi

\bibitem[{{Baldi} {et~al.}(2015){Baldi}, {Capetti}, \&
  {Giovannini}}]{baldi2015}
{Baldi}, R.~D., {Capetti}, A., \& {Giovannini}, G. 2015, \aap, 576, A38

\bibitem[{{Best} {et~al.}(2005){Best}, {Kauffmann}, {Heckman}, {Brinchmann},
  {Charlot}, {Ivezi{\'c}}, \& {White}}]{best2005}
{Best}, P.~N., {Kauffmann}, G., {Heckman}, T.~M., {et~al.} 2005, \mnras, 362,
  25

\bibitem[{{Frank} {et~al.}(2016){Frank}, {Morganti}, {Oosterloo}, {Nyland}, \&
  {Serra}}]{frank2016}
{Frank}, B.~S., {Morganti}, R., {Oosterloo}, T., {Nyland}, K., \& {Serra}, P.
  2016, \aap, 592, A94

\bibitem[{{Intema} {et~al.}(2016){Intema}, {Jagannathan}, {Mooley}, \&
  {Frail}}]{intema2016}
{Intema}, H.~T., {Jagannathan}, P., {Mooley}, K.~P., \& {Frail}, D.~A. 2016,
  ArXiv e-prints

\bibitem[{{Krajnovi{\'c}} {et~al.}(2011){Krajnovi{\'c}}, {Emsellem},
  {Cappellari}, {Alatalo}, {Blitz}, {Bois}, {Bournaud}, {Bureau}, {Davies},
  {Davis}, {de Zeeuw}, {Khochfar}, {Kuntschner}, {Lablanche}, {McDermid},
  {Morganti}, {Naab}, {Oosterloo}, {Sarzi}, {Scott}, {Serra}, {Weijmans}, \&
  {Young}}]{krajnovic2011}
{Krajnovi{\'c}}, D., {Emsellem}, E., {Cappellari}, M., {et~al.} 2011, \mnras,
  414, 2923

\bibitem[{{Laing} \& {Bridle}(2014)}]{laing2014}
{Laing}, R.~A. \& {Bridle}, A.~H. 2014, \mnras, 437, 3405

\bibitem[{{Mauch} \& {Sadler}(2007)}]{mauch2007}
{Mauch}, T. \& {Sadler}, E.~M. 2007, \mnras, 375, 931

\bibitem[{{Morganti} {et~al.}(1997){Morganti}, {Oosterloo}, {Reynolds},
  {Tadhunter}, \& {Migenes}}]{morganti1997}
{Morganti}, R., {Oosterloo}, T.~A., {Reynolds}, J.~E., {Tadhunter}, C.~N., \&
  {Migenes}, V. 1997, \mnras, 284, 541

\bibitem[{{Offringa} {et~al.}(2010){Offringa}, {de Bruyn}, {Biehl}, {Zaroubi},
  {Bernardi}, \& {Pandey}}]{offringa2010}
{Offringa}, A.~R., {de Bruyn}, A.~G., {Biehl}, M., {et~al.} 2010, \mnras, 405,
  155

\bibitem[{{Offringa} {et~al.}(2014){Offringa}, {McKinley}, {Hurley-Walker},
  {Briggs}, {Wayth}, {Kaplan}, {Bell}, {Feng}, {Neben}, {Hughes}, {Rhee},
  {Murphy}, {Bhat}, {Bernardi}, {Bowman}, {Cappallo}, {Corey}, {Deshpande},
  {Emrich}, {Ewall-Wice}, {Gaensler}, {Goeke}, {Greenhill}, {Hazelton},
  {Hindson}, {Johnston-Hollitt}, {Jacobs}, {Kasper}, {Kratzenberg}, {Lenc},
  {Lonsdale}, {Lynch}, {McWhirter}, {Mitchell}, {Morales}, {Morgan},
  {Kudryavtseva}, {Oberoi}, {Ord}, {Pindor}, {Procopio}, {Prabu}, {Riding},
  {Roshi}, {Shankar}, {Srivani}, {Subrahmanyan}, {Tingay}, {Waterson},
  {Webster}, {Whitney}, {Williams}, \& {Williams}}]{offringa2014}
{Offringa}, A.~R., {McKinley}, B., {Hurley-Walker}, N., {et~al.} 2014, \mnras,
  444, 606

\bibitem[{{Offringa} {et~al.}(2012){Offringa}, {van de Gronde}, \&
  {Roerdink}}]{offringa2012}
{Offringa}, A.~R., {van de Gronde}, J.~J., \& {Roerdink}, J.~B.~T.~M. 2012,
  \aap, 539, A95

\bibitem[{{Pandey} {et~al.}(2009){Pandey}, {van Zwieten}, {de Bruyn}, \&
  {Nijboer}}]{pandey2009}
{Pandey}, V.~N., {van Zwieten}, J.~E., {de Bruyn}, A.~G., \& {Nijboer}, R.
  2009, in Astronomical Society of the Pacific Conference Series, Vol. 407, The
  Low-Frequency Radio Universe, ed. D.~J. {Saikia}, D.~A. {Green}, Y.~{Gupta},
  \& T.~{Venturi}, 384

\bibitem[{{S{\'a}nchez-Gallego} {et~al.}(2012){S{\'a}nchez-Gallego}, {Knapen},
  {Wilson}, {Barmby}, {Azimlu}, \& {Courteau}}]{sanchez-gallego2012}
{S{\'a}nchez-Gallego}, J.~R., {Knapen}, J.~H., {Wilson}, C.~D., {et~al.} 2012,
  \mnras, 422, 3208

\bibitem[{{Scaife} \& {Heald}(2012)}]{scaife2012}
{Scaife}, A.~M.~M. \& {Heald}, G.~H. 2012, \mnras, 423, L30

\bibitem[{{Serra} {et~al.}(2012){Serra}, {Oosterloo}, {Morganti}, {Alatalo},
  {Blitz}, {Bois}, {Bournaud}, {Bureau}, {Cappellari}, {Crocker}, {Davies},
  {Davis}, {de Zeeuw}, {Duc}, {Emsellem}, {Khochfar}, {Krajnovi{\'c}},
  {Kuntschner}, {Lablanche}, {McDermid}, {Naab}, {Sarzi}, {Scott}, {Trager},
  {Weijmans}, \& {Young}}]{serra2012}
{Serra}, P., {Oosterloo}, T., {Morganti}, R., {et~al.} 2012, \mnras, 422, 1835

\bibitem[{{van Weeren} {et~al.}(2016){van Weeren}, {Williams}, {Hardcastle},
  {Shimwell}, {Rafferty}, {Sabater}, {Heald}, {Sridhar}, {Dijkema}, {Brunetti},
  {Br{\"u}ggen}, {Andrade-Santos}, {Ogrean}, {R{\"o}ttgering}, {Dawson},
  {Forman}, {de Gasperin}, {Jones}, {Miley}, {Rudnick}, {Sarazin}, {Bonafede},
  {Best}, {B{\^i}rzan}, {Cassano}, {Chy{\.z}y}, {Croston}, {Ensslin},
  {Ferrari}, {Hoeft}, {Horellou}, {Jarvis}, {Kraft}, {Mevius}, {Intema},
  {Murray}, {Orr{\'u}}, {Pizzo}, {Simionescu}, {Stroe}, {van der Tol}, \&
  {White}}]{vanWeeren2016}
{van Weeren}, R.~J., {Williams}, W.~L., {Hardcastle}, M.~J., {et~al.} 2016,
  \apjs, 223, 2

\bibitem[{{Williams} {et~al.}(2016){Williams}, {van Weeren}, {R{\"o}ttgering},
  {Best}, {Dijkema}, {de Gasperin}, {Hardcastle}, {Heald}, {Prandoni},
  {Sabater}, {Shimwell}, {Tasse}, {van Bemmel}, {Br{\"u}ggen}, {Brunetti},
  {Conway}, {En{\ss}lin}, {Engels}, {Falcke}, {Ferrari}, {Haverkorn},
  {Jackson}, {Jarvis}, {Kapi{\'n}ska}, {Mahony}, {Miley}, {Morabito},
  {Morganti}, {Orr{\'u}}, {Retana-Montenegro}, {Sridhar}, {Toribio}, {White},
  {Wise}, \& {Zwart}}]{williams2016}
{Williams}, W.~L., {van Weeren}, R.~J., {R{\"o}ttgering}, H.~J.~A., {et~al.}
  2016, \mnras

\end{thebibliography}


\begin{thebibliography}{}

\bibitem[Adams \& van Leeuwen(2019)]{Adams19} Adams, E.~A.~K. \& van Leeuwen, J.\ 2019, Nature Astronomy, 3, 188

\bibitem[Barthel et al.(1985)]{Barthel85} Barthel, P.~D., Schilizzi, R.~T., Miley, G.~K., et al.\ 1985, \aap, 148, 243

\bibitem[Brienza et al.(2016)]{Brienza16} Brienza, M., Godfrey, L., Morganti, R., et al.\ 2016, \aap, 585, A29

\bibitem[Brienza et al.(2017)]{Brienza17} Brienza, M., Godfrey, L., Morganti, R., et al.\ 2017, \aap, 606, A98

\bibitem[Brienza et al.(2020)]{Brienza20} Brienza, M., Morganti, R., Harwood, J., et al.\ 2020, \aap, 638, A29

\bibitem[Briggs (1995)]{Briggs95} Briggs D., 1995, PhD Thesis, New Mexico Institude of Mining and Technology (http://www.aoc.nrao.edu/dissertations/dbriggs/)

\bibitem[Brocksopp et al.(2007)]{Brocksopp07} Brocksopp, C., Kaiser, C.~R., Schoenmakers, A.~P., et al.\ 2007, \mnras, 382, 1019

\bibitem[Bruni et al.(2020)]{Bruni20} Bruni, G., Panessa, F., Bassani, L., et al.\ 2020, \mnras, 494, 902

\bibitem[Callingham et al.(2017)]{Callingham17} Callingham, J.~R., Ekers, R.~D., Gaensler, B.~M., et al.\ 2017, \apj, 836, 174 

\bibitem[Ciotti et al.(2010)]{Ciotti10} Ciotti, L., Ostriker, J.~P., \& Proga, D.\ 2010, \apj, 717, 708

\bibitem[Coppejans et al.(2017)]{Coppejans17} Coppejans, R., van Velzen, S., Intema, H.~T., et al.\ 2017, \mnras, 467, 2039

\bibitem[Cordey(1987)]{Cordey87} Cordey, R.~A.\ 1987, \mnras, 227, 695

\bibitem[Croston et al.(2008)]{Croston08} Croston, J.~H., Hardcastle, M.~J., Birkinshaw, M., et al.\ 2008, \mnras, 386, 1709


\bibitem[Duncan et al.(2019)]{Duncan19} Duncan, K.~J., Sabater, J., R{\"o}ttgering, H.~J.~A., et al.\ 2019, \aap, 622, A3

\bibitem[Duncan et al.(2020)]{Duncan20} Duncan et al. 2020, A\&A, submitted (this issue)

\bibitem[Eilek(1996)]{Eilek96} Eilek, J.~A.\ 1996, Energy Transport in Radio Galaxies and Quasars, 281

\bibitem[English et al.(2019)]{English19} English, W., Hardcastle, M.~J., \& Krause, M.~G.~H.\ 2019, \mnras, 490, 5807

\bibitem[Fanaroff, \& Riley(1974)]{Fanaroff74} Fanaroff, B.~L., \& Riley, J.~M.\ 1974, MNRAS, 167, 31P

\bibitem[Gabor \& Bournaud(2013)]{Gabor13} Gabor, J.~M., \& Bournaud, F.\ 2013, \mnras, 434, 606

\bibitem[Galvin et al.(2019)]{Galvin19} Galvin, T.~J., Huynh, M., Norris, R.~P., et al.\ 2019, \pasp, 131, 108009

\bibitem[Gaspari et al.(2017)]{Gaspari17} Gaspari, M., Temi, P., \& Brighenti, F.\ 2017, \mnras, 466, 677

\bibitem[de Gasperin et al.(2014)]{Gasperin14} de Gasperin, F., Intema, H.~T., Williams, W., et al.\ 2014, \mnras, 440, 1542

\bibitem[de Gasperin et al.(2018)]{Gasperin18} de Gasperin, F., Intema, H.~T., \& Frail, D.~A.\ 2018, \mnras, 474, 5008


\bibitem[Gizani \& Leahy(2003)]{Gizani03} Gizani, N.~A.~B., \& Leahy, J.~P.\ 2003, \mnras, 342, 399

\bibitem[Godfrey et al.(2017)]{Godfrey17} Godfrey, L.~E.~H., Morganti, R., \& Brienza, M.\ 2017, \mnras, 471, 891

\bibitem[van Haarlem et al.(2013)]{Haarlem13} van Haarlem, M.~P., Wise, M.~W., Gunst, A.~W., et al.\ 2013, \aap, 556, A2

\bibitem[Hardcastle(2018)]{Hardcastle18} Hardcastle, M.~J.\ 2018, \mnras, 475, 2768

\bibitem[Hardcastle \& Croston(2020)]{Hardcastle20} Hardcastle, M.~J., \& Croston, J.~H.\ 2020, arXiv e-prints, arXiv:2003.06137

\bibitem[Harwood et al.(2016)]{Harwood16} Harwood, J.~J., Croston, J.~H., Intema, H.~T., et al.\ 2016, \mnras, 458, 4443

\bibitem[Heckman \& Best(2014)]{Heckman14} Heckman, T.~M. \& Best, P.~N.\ 2014, \araa, 52, 589

\bibitem[Heesen et al.(2018)]{Heesen18} Heesen, V., Croston, J.~H., Morganti, R., et al.\ 2018, \mnras, 474, 5049

\bibitem[Hobbs et al.(2011)]{Hobbs11} Hobbs, A., Nayakshin, S., Power, C., et al.\ 2011, \mnras, 413, 2633

\bibitem[Hurley-Walker et al.(2015)]{Hurley15} Hurley-Walker, N., Johnston-Hollitt, M., Ekers, R., et al.\ 2015, \mnras, 447, 2468

\bibitem[Hurley-Walker et al.(2017)]{Hurley17} Hurley-Walker, N., Callingham, J.~R., Hancock, P.~J., et al.\ 2017, \mnras, 464, 1146

\bibitem[Ineson et al.(2017)]{Ineson17} Ineson, J., Croston, J.~H., Hardcastle, M.~J., et al.\ 2017, \mnras, 467, 1586

\bibitem[Intema et al.(2011)]{Intema11} Intema, H.~T., van Weeren, R.~J., R{\"o}ttgering, H.~J.~A., et al.\ 2011, \aap, 535, A38

\bibitem[Intema et al.(2017)]{Intema17} Intema, H.~T., Jagannathan, P., Mooley, K.~P., et al.\ 2017, \aap, 598, A78

\bibitem[Kardashev(1962)]{Kardashev62} Kardashev, N.~S.\ 1962, \sovast, 6, 317

\bibitem[Kellermann(1964)]{Kellerman64} Kellermann, K.~I.\ 1964, \apj, 140, 969

\bibitem[Klamer et al.(2006)]{Klamer06} Klamer, I.~J., Ekers, R.~D., Bryant, J.~J., et al.\ 2006, \mnras, 371, 852

\bibitem[Komissarov \& Gubanov(1994)]{Komissarov94} Komissarov, S.~S., \& Gubanov, A.~G.\ 1994, \aap, 285, 27

\bibitem[Konar et al.(2013)]{Konar13} Konar, C., Hardcastle, M.~J., Jamrozy, M., et al.\ 2013, \mnras, 430, 2137

\bibitem[Kondapally et al.(2020)]{Kondapally20} Kondapally et al. 2020, A\&A, submitted (this issue)

\bibitem[Ku{\'z}micz et al.(2017)]{Kuzmicz17} Ku{\'z}micz, A., Jamrozy, M., Kozie{\l}-Wierzbowska, D., et al.\ 2017, \mnras, 471, 3806

\bibitem[Jaffe, \& Perola(1973)]{Jaffe73} Jaffe, W.~J., \& Perola, G.~C.\ 1973, \aap, 26, 423

\bibitem[Jamrozy et al.(2007)]{Jamrozy07} Jamrozy, M., Konar, C., Saikia, D.~J., et al.\ 2007, \mnras, 378, 581

\bibitem[Joshi et al.(2011)]{Joshi11} Joshi, S.~A., Nandi, S., Saikia, D.~J., et al.\ 2011, \mnras, 414, 1397

\bibitem[Jurlin et al.(2020)]{Jurlin20} Jurlin, N., Morganti, R., Brienza, M., et al.\ 2020, A\&A in press, arXiv e-prints, arXiv:2004.09118

\bibitem[Maccagni et al.(2020)]{Maccagni20} Maccagni, F.~M., Murgia, M., Serra, P., et al.\ 2020, \aap, 634, A9

\bibitem[Mahatma et al.(2018)]{Mahatma18} Mahatma, V.~H., Hardcastle, M.~J., Williams, W.~L., et al.\ 2018, \mnras, 475, 4557

\bibitem[Mahatma et al.(2019)]{Mahatma19} Mahatma, V.~H., Hardcastle, M.~J., Williams, W.~L., et al.\ 2019, \aap, 622, A13

\bibitem[Mahony et al.(2016)]{Mahony16} Mahony, E.~K., Morganti, R., Prandoni, I., et al.\ 2016, \mnras, 463, 2997

\bibitem[McNamara, \& Nulsen(2012)]{McNamara12} McNamara, B.~R., \& Nulsen, P.~E.~J.\ 2012, New Journal of Physics, 14, 055023

\bibitem[Miley \& De Breuck(2008)]{Miley08} Miley, G., \& De Breuck, C.\ 2008, \aapr, 15, 67

\bibitem[Mingo et al.(2019)]{Mingo19} Mingo, B., Croston, J.~H., Hardcastle, M.~J., et al.\ 2019, \mnras, 488, 2701

\bibitem[Mohan \& Rafferty(2015)]{Mohan15} Mohan, N., \& Rafferty, D.\ 2015, PyBDSF: Python Blob Detection and Source Finder, ascl:1502.007

\bibitem[Morabito \& Harwood(2018)]{Morabito18} Morabito, L.~K., \& Harwood, J.~J.\ 2018, \mnras, 480, 2726

\bibitem[Morganti et al.(2011)]{Morganti11} Morganti, R., Holt, J., Tadhunter, C., et al.\ 2011, \aap, 535, A97

\bibitem[Morganti(2017)]{Morganti17} Morganti, R.\ 2017, Nature Astronomy, 1, 596

\bibitem[Morganti \& Oosterloo(2018)]{Morganti18} Morganti, R., \& Oosterloo, T.\ 2018, \aapr, 26, 4

\bibitem[Murgia et al.(2011)]{Murgia11} Murgia, M., Parma, P., Mack, K.-H., et al.\ 2011, \aap, 526, A148

\bibitem[Ralph et al.(2019)]{Ralph19} Ralph, N.~O., Norris, R.~P., Fang, G., et al.\ 2019, \pasp, 131, 108011

\bibitem[Novak et al.(2011)]{Novak11} Novak, G.~S., Ostriker, J.~P., \& Ciotti, L.\ 2011, \apj, 737, 26

\bibitem[O'Dea(1998)]{ODea98} O'Dea, C.~P.\ 1998, \pasp, 110, 493

\bibitem[Offringa et al.(2012)]{Offringa12} Offringa, A.~R., van de Gronde, J.~J., \& Roerdink, J.~B.~T.~M.\ 2012, \aap, 539, A95

\bibitem[Offringa et al.(2010)]{Offringa10} Offringa, A.~R., de Bruyn, A.~G., Biehl, M., et al.\ 2010, \mnras, 405, 155


\bibitem[Orr{\`u} et al.(2010)]{Orru10} Orr{\`u}, E., Murgia, M., Feretti, L., et al.\ 2010, \aap, 515, A50

\bibitem[Oosterloo et al.(2007)]{Oosterloo07} Oosterloo, T., Fraternali, F., \& Sancisi, R.\ 2007, \aj, 134, 1019

\bibitem[Oosterloo et al.(2018)]{Oosterloo18} Oosterloo, T., van Leeuwen, J., Van Cappellen, W., et al.\ 2018, Westerbork Telescope 50th Anniversary, 16

\bibitem[Orienti(2016)]{Orienti16} Orienti, M.\ 2016, Astronomische Nachrichten, 337, 9

\bibitem[Orr{\`u} et al.(2015)]{Orru15} Orr{\`u}, E., van Velzen, S., Pizzo, R.~F., et al.\ 2015, \aap, 584, A112

\bibitem[Pacholczyk(1970)]{Pacholczyk70} Pacholczyk, A.~G.\ 1970, Radio astrophysics: Nonthermal processes in galactic and extragalactic sources. Phys. Today 24

\bibitem[Parma et al.(2007)]{Parma07} Parma, P., Murgia, M., de Ruiter, H.~R., et al.\ 2007, \aap, 470, 875

\bibitem[Parma et al.(2010)]{Parma10} Parma, P., Mantovani, F., de Ruiter, H.~R., et al.\ 2010, 10th European VLBI Network Symposium and EVN Users Meeting: VLBI and the New Generation of Radio Arrays, 89


\bibitem[Prandoni et al.(2018)]{Prandoni18} Prandoni, I., Guglielmino, G., Morganti, R., et al.\ 2018, \mnras, 481, 4548

\bibitem[Roettiger et al.(1994)]{Roettiger94} Roettiger, K., Burns, J.~O., Clarke, D.~A., et al.\ 1994, \apjl, 421, L23

\bibitem[Sabater et al.(2019)]{Sabater19} Sabater, J., Best, P.~N., Hardcastle, M.~J., et al.\ 2019, \aap, 622, A17

\bibitem[Sabater et al.(2020)]{Sabater20} Sabater et al. 2020, A\&A, submitted (this issue)

\bibitem[Saikia, \& Jamrozy(2009)]{Saikia09} Saikia, D.~J., \& Jamrozy, M.\ 2009, Bulletin of the Astronomical Society of India, 37, 63

\bibitem[Saripalli et al.(2012)]{Saripalli12} Saripalli, L., Subrahmanyan, R., Thorat, K., et al.\ 2012, \apjs, 199, 27

\bibitem[Sault et al.(1995)]{Sault95} Sault, R.~J., Teuben, P.~J., \& Wright, M.~C.~H.\ 1995, Astronomical Data Analysis Software and Systems IV, 77, 433 

\bibitem[Schulz et al.(2019)]{Schulz19} Schulz, R. et al. 2019 Proceedings of the ADASS Conference 

\bibitem[Shabala et al.(2020)]{Shabala20} Shabala, S., Jurlin, N., Morganti, R., et al.\ 2020, MNRAS in press, arXiv e-prints, arXiv:2004.08979

\bibitem[Shimwell et al.(2019)]{Shimwell19} Shimwell, T.~W., Tasse, C., Hardcastle, M.~J., et al.\ 2019, \aap, 622, A1

\bibitem[Schoenmakers et al.(2000)]{Schoenmakers00} Schoenmakers, A.~P., de Bruyn, A.~G., R{\"o}ttgering, H.~J.~A., et al.\ 2000, \mnras, 315, 371
\bibitem[Shulevski et al.(2012)]{Shulevski12} Shulevski, A., Morganti, R., Oosterloo, T., et al.\ 2012, \aap, 545, A91

\bibitem[Shulevski et al.(2015)]{Shulevski15} Shulevski, A., Morganti, R., Barthel, P.~D., et al.\ 2015, \aap, 579, A27

\bibitem[Shulevski et al.(2017)]{Shulevski17} Shulevski, A., Morganti, R., Harwood, J.~J., et al.\ 2017, \aap, 600, A65

\bibitem[Tamhane et al.(2015)]{Tamhane15} Tamhane, P., Wadadekar, Y., Basu, A., et al.\ 2015, \mnras, 453, 2438

\bibitem[Tasse et al.(2020)]{Tasse20} Tasse et al. 2020, A\&A, submitted (this issue)

\bibitem[Turner \& Shabala(2015)]{Turner15} Turner, R.~J., \& Shabala, S.~S.\ 2015, \apj, 806, 59

\bibitem[Turner et al.(2018)]{Turner18} Turner, R.~J., Rogers, J.~G., Shabala, S.~S., et al.\ 2018, \mnras, 473, 4179

\bibitem[Turner \& Shabala(2020)]{Turner20} Turner, R.~J., \& Shabala, S.~S.\ 2020, \mnras, 493, 5181

\bibitem[Vantyghem et al.(2014)]{Vantyghem14} Vantyghem, A.~N., McNamara, B.~R., Russell, H.~R., et al.\ 2014, \mnras, 442, 3192

\bibitem[van Weeren et al.(2014)]{Weeren14} van Weeren, R.~J., Williams, W.~L., Tasse, C., et al.\ 2014, \apj, 793, 82


\bibitem[Wilber et al.(2018)]{Wilber18} Wilber, A., Br{\"u}ggen, M., Bonafede, A., et al.\ 2018, \mnras, 473, 3536

\bibitem[Williams et al.(2013)]{Williams13} Williams, W.~L., Intema, H.~T., \& R{\"o}ttgering, H.~J.~A.\ 2013, \aap, 549, A55

\bibitem[Williams et al.(2016)]{Williams16} Williams, W.~L., van Weeren, R.~J., R{\"o}ttgering, H.~J.~A., et al.\ 2016, \mnras, 460, 2385

\end{thebibliography}

\newpage


\newpage

\begin{appendix} 
        \section{Core-dominated sources}
        \label{sec:appendix1}

\begin{figure*}[h]
   \centering
  \includegraphics[width=4.5cm,angle=0]{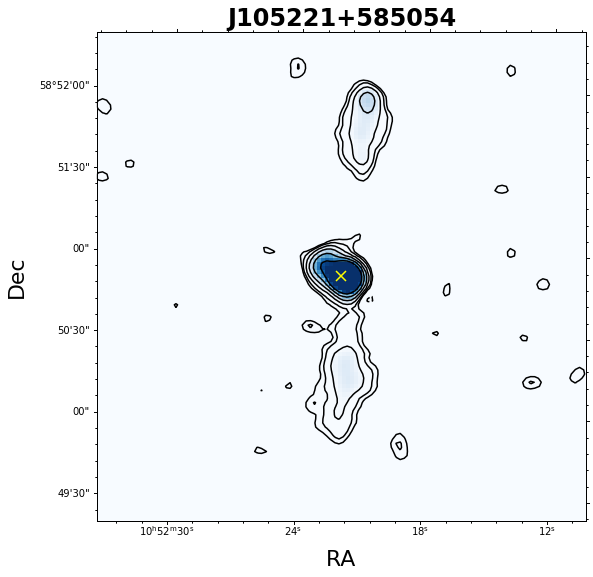} 
   \includegraphics[width=4.5cm,angle=0]{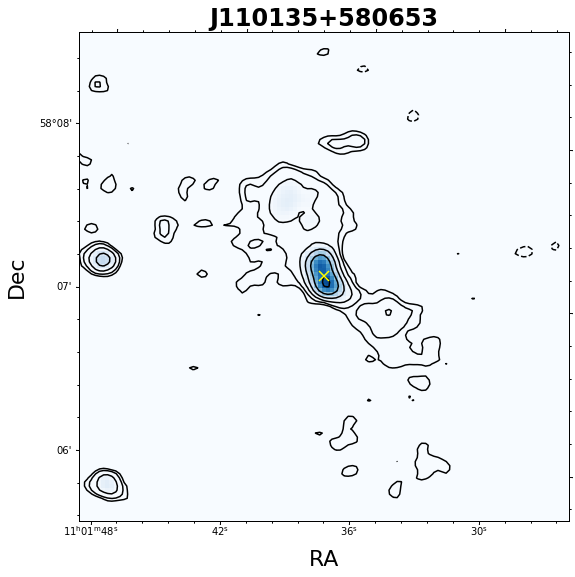} 
   \includegraphics[width=4.5cm,angle=0]{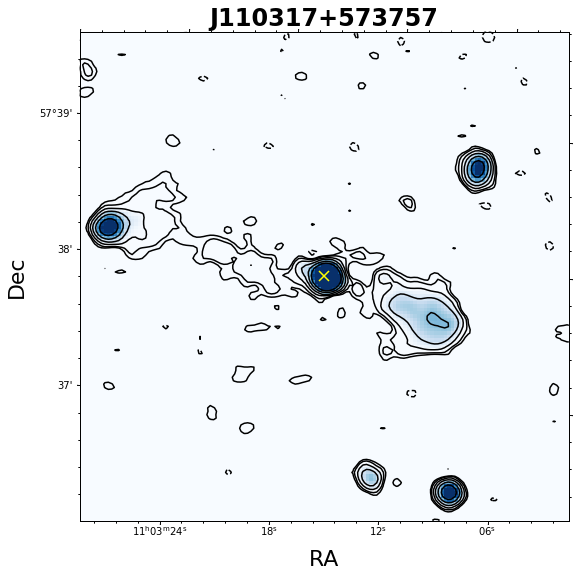} 
\caption{LOFAR 6 arcsec cutout of some of the sources shown in Fig. \ref{fig:partUSS}. Contour levels are $-3, 3, 5, 10, 20, 40,$ and $80 \times \sigma$, where $\sigma = 50$ \muJybeam. From left to right: J105221+58, J110135+58, and J110317+57. The position of the core (and optical identification) is marked by the cross.}
\label{fig:partUSS_HR}
\end{figure*}

\section{Abell~1132}
\label{sec:appendix2}
        
\begin{figure*}[h]
   \centering
\includegraphics[width=9.5cm,angle=-90]{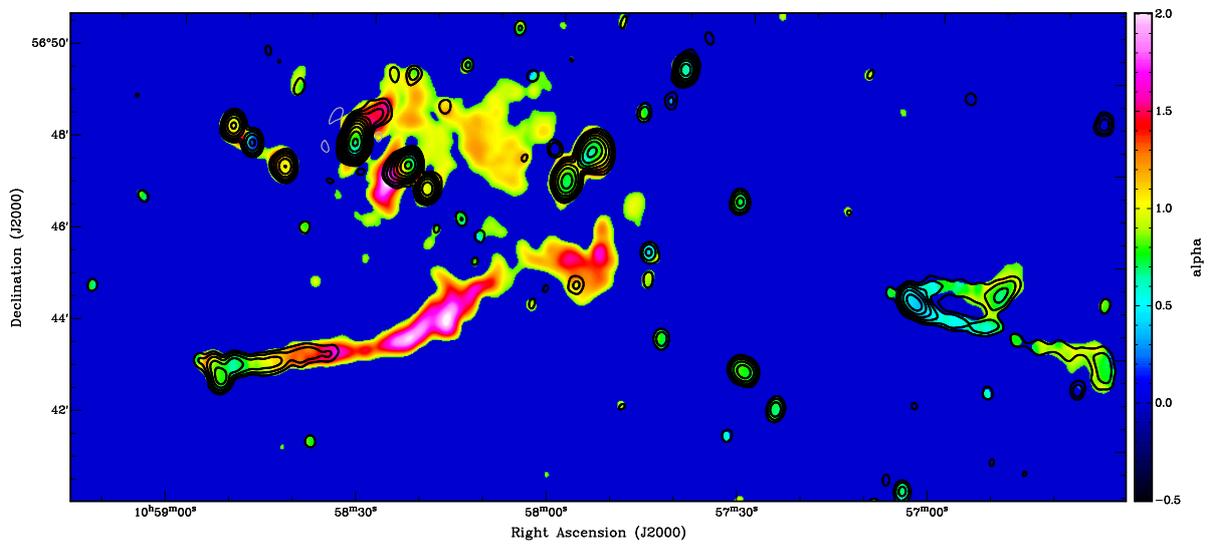}
   \caption{Images of the spectral index (\si) and spectral index limits for the region of Abell~1132.  The values of the spectral index in the initial part of the tail are consistent with what is presented by \cite{Wilber18}, and the sharp steepenings in the outer part are clearly seen. The contours are from the Apertif image with logarithmic levels starting at  0.15 \mJybeam\ and increasing with a factor 2.
}
\label{fig:A1132}
\end{figure*}

\end{appendix}

\end{document}